\documentclass{article}

\usepackage{PRIMEarxiv}

\usepackage[utf8]{inputenc} % allow utf-8 input
\usepackage[T1]{fontenc}    % use 8-bit T1 fonts
\usepackage{hyperref}       % hyperlinks
\usepackage{url}            % simple URL typesetting
\usepackage{booktabs}       % professional-quality tables
\usepackage{amsfonts}       % blackboard math symbols
\usepackage{nicefrac}       % compact symbols for 1/2, etc.
\usepackage{microtype}      % microtypography
\usepackage{lipsum}
\usepackage{fancyhdr}       % header
\usepackage{graphicx}       % graphics
\usepackage{savesym}
\usepackage{amsmath}
\usepackage{cases} 
\usepackage{textcomp}
\usepackage{xspace}
\usepackage{verbatim}
\usepackage{soul}
\usepackage{xcolor}
\usepackage{blindtext}
\usepackage{verbatim}
\usepackage{algorithm}
\savesymbol{AND}    
\usepackage{algorithmic}
\usepackage{tabularx} 
\usepackage{enumerate}
\usepackage{multirow}
\usepackage{diagbox}
\usepackage{colortbl}
\usepackage{hhline}
\usepackage{comment}
\usepackage{balance}

\usepackage{dsfont}
\usepackage{wrapfig}
\usepackage[defaultlines=4,all]{nowidow}
\definecolor{darkblue}{rgb}{0,0,0.5} 
\usepackage{transparent}



\makeatother

\newcommand{\hybrid}{HCR\xspace}

\newcommand{\vbools}{VBools\xspace}

\newcommand{\additiveS}{\textit{Additive}\xspace}
\newcommand{\maxS}{\textit{Max}\xspace}
\newcommand{\CPSs}{CPSs\xspace}

\newcommand{\BO}{BO\xspace}
\newcommand{\TR}{TR\xspace}
\newcommand{\LineSearch}{LSF\xspace}
\newcommand{\REMBO}{REMBO\xspace}

\newcommand{\VanillaBO}{vanilla BO\xspace}
\newcommand{\turbo}{TuRBO\xspace}

\newcommand{\SUT}{SUT\xspace}

\newcommand{\PI}{PI\xspace}

\newcommand{\LCB}{LCB\xspace}

\newcommand{\TS}{TS\xspace}
\newcommand{\STL}{STL\xspace}
\newcommand{\ATprime}[1]{\ensuremath{\varphi_#1^{AT^\prime}}\xspace}
\newcommand{\tauZero}[1]{\ensuremath{\tau = 0}\xspace} 
\newcommand{\tauOne}[1]{\ensuremath{\tau = -1}\xspace} 
\newcommand{\piBO}[1]{\ensuremath{\pi}BO\xspace}

\usepackage{todonotes}

\newcommand{\AT}{{\it AT}\xspace}

\newcommand{\SC}{{\it SC}\xspace}

\DeclareMathOperator{\lnotvb}{\lnot_v}

\DeclareUnicodeCharacter{2212}{-}

\newcommand{\xtrace}{x^s\xspace}
\newcommand{\xtracein}{x^s_i\xspace}
\newcommand{\ytraceout}{x^s_o\xspace}
\title{Falsification of Cyber-Physical Systems using Bayesian Optimization
}

\author{
  Zahra~Ramezani \\
  Chalmers University of Technology \\
  Gothenburg, Sweden \\
  \texttt{rzahra@chalmers.se} \\
   \And
  Kenan~\v{S}ehi\'c \\
  Lund University, Lund, Sweden,
  \\University of Sarajevo, Bosnia and Herzegovina \\
  \texttt{kenan.sehic@cs.lth.se, ksehic@etf.unsa.ba} \\
  \AND
  Luigi Nardi \\
  Lund University and Stanford University \\
  DBtune, Sweden \\
  \texttt{luigi.nardi@cs.lth.se, lnardi@stanford.edu} \\
  \And
  Knut {\AA}kesson \\
  Chalmers University of Technology \\
  Gothenburg, Sweden \\
  \texttt{knut@chalmers.se} \\
}

\begin{document}
\maketitle

\begin{abstract}
Cyber-physical systems (CPSs) are often complex and safety-critical, making it both challenging and crucial to ensure that the system's specifications are met. Simulation-based falsification is a practical testing technique for increasing confidence in a CPS's correctness, as it only requires that the system be simulated. Reducing the number of computationally intensive simulations needed for falsification is a key concern. In this study, we investigate Bayesian optimization (BO), a sample-efficient approach that learns a surrogate model to capture the relationship between input signal parameterization and specification evaluation. We propose two enhancements to the basic BO for improving falsification: (1) leveraging local surrogate models, and (2) utilizing the user's prior knowledge. Additionally, we address the formulation of acquisition functions for falsification by proposing and evaluating various alternatives. Our benchmark evaluation demonstrates significant improvements when using local surrogate models in BO for falsifying challenging benchmark examples. Incorporating prior knowledge is found to be especially beneficial when the simulation budget is constrained. For some benchmark problems, the choice of acquisition function noticeably impacts the number of simulations required for successful falsification.

\end{abstract}

% keywords can be removed

% Introduction
%%%%% Introduction:
\section{Introduction}
\label{Sec:Introduction}
 
% Cyber-Physical Systems:
% Challenges for simulation-based falsification using optimization:
%Falsification is a requirements-based testing approach that searches for input traces to the system that falsify given requirements.

Cyber-physical systems (CPS)~\cite{alur2015principles} are frequently complex and safety-critical. Testing is a common approach for evaluating correctness, with the primary goal of identifying inputs that falsify given specifications, known as \emph{counterexamples}.

In many industrial systems, explicit mathematical models for analysis are unavailable, and only system simulations are possible. In this paper, we assume the availability of a black-box model representing the system under test (\SUT), which can be simulated.

For numerous industrial CPSs, simulations can be computationally expensive, making it desirable to minimize the number of simulations required during falsification. \emph{Simulation-based falsification using optimization} aims to reduce the number of tests, i.e., simulations, by employing an optimization method to determine the next set of input signals based on previous simulation results. A key challenge lies in selecting an efficient optimization method and identifying the information an optimizer should use to decide on the next input signals.

When the \SUT is represented as a black-box model, the optimization approach is limited to gradient-free optimization methods~\cite{audet2017}. Optimization-based methods are generally divided into two categories: \emph{direct-search}~\cite{Hooke1961direct} and \emph{model-based} methods~\cite{audet2017}. Direct-search methods evaluate the objective function directly without sharing information between consecutive evaluations, while model-based searches create a surrogate model of the objective function to explore and exploit the search space~\cite{audet2017,Shahriari2016}.

Model-based methods are suitable when the \SUT is expensive to simulate, and it might be worth the additional cost of building a surrogate model. Bayesian optimization (\BO)~\cite{Shahriari2016} is a model-based optimization method that has been successfully applied to various problems, such as machine learning hyperparameter optimization~\cite{hvarfner2022pi,vsehic2021lassobench} and robotics~\cite{mayr2022skill,berkenkamp2021bayesian}.
%Often, BO outperforms conventional optimization techniques when addressing costly-to-evaluate nonconvex functions with multiple local optima.

Often \BO outperforms conventional optimization when addressing costly-to-evaluate nonconvex functions with multiple local optima~\cite{Shahriari2016}. As falsification of \CPSs is typically a high-dimensional problem, the computational requirements for vanilla \BO (i.e., the standard Bayesian optimization approach) are not well suited.

%in these settings exponentially increase and easily become impractical. 
%Furthermore, exploring and exploiting the input space becomes ambiguous where the edges, an upper and/or lower bound of input ranges, are more frequently exploited. 
In~\cite{Deshmukh2017Bayesian}, it is demonstrated that \BO is comparable with respect to other optimization methods, such as CMA-ES~\cite{hansen2006cma}, for falsification of \CPSs. However,  \cite{Deshmukh2017Bayesian} is using a dimensionality reduction method \REMBO~\cite{Wang2016REMBO}. \REMBO and its extensions, such as HeSBO~\cite{nayebi2019} and ALEBO~\cite{alebo}, project the search space to a low-dimensional subspace using a linear embedding. In these methods, a Gaussian process model is trained on a low-dimensional space, from which the original high-dimensional input space for evaluations is derived via inverse random projection. Consequently, points outside the search domain are projected to the boundary, upper or lower bound, resulting in over-exploitation of the boundary region. However, determining the optimal size of a linear embedding is a challenging task in practice.
Creating an adequate linear embedding with the optimal size for different falsification problems is an open research question. In general, it is impractical for practitioners to specify linear embeddings in real-world applications without having multiple intensive trials.

Moreover, in~\cite{Deshmukh2017Bayesian}, the authors use different linear embeddings for each example, and there is no clear suggestion on which embedding is optimal. Furthermore, the benchmark problems used in ~\cite{Deshmukh2017Bayesian} have been updated to more recent benchmarking problems used in the falsification community~\cite{Ernst2019-ARCH19:, Ramezani2020-wodes}.

%, as shown in~\cite{Deshmukh2017Bayesian}.  

% Local search, turbo
A different and more practical approach would be to use an approach that does not require input from the practitioner. \turbo~\cite{Eriksson2019turbo} is such a method that is based on trust regions~\cite{Yuan2000ReviewTrust}. A trust region (TR) is a subset of the input space centered at the current-best solution where the objective function is approximated locally. Instead of focusing on a linear embedding, \turbo searches for the objective function locally with a sequence of local optimization runs. Using TRs in which a probabilistic model is trained similarly to the global \BO framework helps to avoid overexploiting and thus provides a balance between exploration and exploitation.
In some practical applications of falsification, the objective function can be constant in large regions %lacks smoothness 
or may have discontinuities. In these situations, vanilla \BO will have difficulty learning anything credible. However, this limitation can be easily omitted by shrinking the search locally within a TR as done in \turbo. Although, the final performance of \BO depends crucially on the selection of the acquisition function. The acquisition function determines how exploration and exploitation are balanced in \BO~\cite{Thompson1933likelihood, Srinivas2010UCB,Kushner1964PI}, an aspect that is not discussed in \cite{Deshmukh2017Bayesian}. %acquisition function how exploration and exploitation of the search space are done in \BO)~\cite{Thompson1933likelihood, Srinivas2010UCB, Kushner1964PI}, the literature~\cite{Deshmukh2017Bayesian} for falsification using \BO mostly neglects }this effect.} 
%Depending on an application, there can be a preference for one acquisition function over another. Moreover, an algorithm's performance heavily depends on how well the acquisition function balances exploration and exploitation. 
One acquisition function can be used to emphasize model uncertainty more than prediction, while another acquisition function can be used to accelerate convergence but might get trapped in local optima easily. To address this issue, in addition to the default choice with Thompson Sampling (\TS)~\cite{Thompson1933likelihood}, \turbo is modified, as part of this work, to use the lower confidence bound (LCB)~\cite{Srinivas2010UCB}, and an adjusted version of the probability of improvement (\PI)~\cite{Kushner1964PI} that emphasizes more failure events as our primary goal is to find configurations that falsify the \SUT. 

% Prior BO, piBO
Despite being a popular method for optimizing expensive black-box functions, plain \BO does not incorporate the expertise of domain experts.
For certain falsification problems, there is prior knowledge about where a falsified point can potentially be located. Such an example is the corner points, that is, values at the boundaries of the allowed input ranges, that are shown to be likely to falsify many falsification problems; this is further discussed in~\cite{Ramezani2021_Line}. In such situations, prior knowledge can be incorporated into the model-based method to increase efficiency. The recent works, BOPrO~\cite{Souza2021Prior} and $\pi$BO~\cite{hvarfner2022pi} propose how to incorporate the prior injection about the optimal solution in \BO, which allows the practitioner to emphasize certain regions that potentially a falsified point can be located. The methods can forget incorrect prior knowledge and eventually converge to an optimal solution. To our knowledge, injecting a prior belief about the falsifiable area has never been evaluated for falsification of \CPSs. In~\cite{Souza2021Prior}, a user-provided prior distribution is combined with a data-driven model to form a pseudo-posterior. Still, this approach does not allow for arbitrary priors to be integrated. Furthermore, this method does not allow for different acquisition functions. 
A generalized approach, $\pi$BO, is proposed in~\cite{hvarfner2022pi} to address these issues. In contrast to other works, while being conceptually simple, $\pi$BO can easily be integrated with existing \BO works and different acquisition functions. Furthermore, ~\cite{hvarfner2022pi} provides a theoretical guarantee proving convergence at regular rates independently of the prior.

In this paper, we investigate how \turbo, a BO method for high-dimensional problems, and $\pi$BO, a BO method for including prior knowledge, can be used for falsification. We propose the modification of the \turbo method with different acquisition functions, evaluate them on benchmark problems, and compare them with other state-of-the-art methods.

% Contributions
These mentioned limitations and shortcomings can be one of the reasons why \BO has not yet received much attention in the falsification community. Moreover, since the initial work~\cite{Deshmukh2017Bayesian} on \BO was proposed a few years ago, the new standard benchmark problems for falsification have emerged~\cite{Ernst2019-ARCH19:,Ramezani2020-wodes}.  
Therefore, a new study on \BO optimization for falsification is necessary. We have formulated the following research questions in this paper. (1) How can recent contributions from the BO community be exploited to efficiently solve falsification problems? (2) Which formulation of the acquisition functions is suitable for the falsification process? (3) How can guesses about falsification points be included in the BO approach?
The contributions in this paper are: (i) A discussion of recently proposed BO optimization methods that include trust regions and allow the possibility to inject prior knowledge in the problem formulation. In particular, we discuss how these new features are affecting the falsification process. (ii) A new acquisition function that is tailor-made for falsification problems is proposed. (iii) An extensive evaluation of the proposed methods and acquisition functions on a representative set of benchmark problems. 
The evaluations demonstrate that recent trust-region-based Bayesian optimization is, for hard problems to falsify, outperforming traditional optimization-based approaches.

%The choice of acquisition function and the inclusion of priors ha

%outperforms the state-of-the-art, including both optimization-free and optimization-based methods, falsification methods for some specifications and examples.}  

% Organization of the paper: 
This paper is organized as follows: Section~\ref{Sec:Related-Work} reviews the related work. Section~\ref{Sec:Backgrould-Falsification-BO} introduces the first falsification process briefly, then the \BO method.
Section~\ref{Sec: Method} introduces the methods used in this paper.
The evaluation results on the benchmark problems are discussed in Section~\ref{Sec:experimental}. Finally, Section~\ref{Sec:Conclusion} summarizes the contributions.

% Related Work
\section{Related Work}
\label{Sec:Related-Work}
% Different evaluated optimization methods for falsification
Different optimization methods have been evaluated for falsification of \CPSs. In~\cite{Eddeland2020-ARCH}, the optimization methods, Covariance Matrix Adaptation Evolution Strategy (CMA-ES)~\cite{hansen2006cma}, Nelder-Mead (NM)~\cite{Nelder1965}, SNOBFIT~\cite{huyer2008snobfit},  and Simulated Annealing~\cite{romeijn1994simulated} were evaluated on benchmark problems~\cite{Ernst2019-ARCH19:} to evaluate the effect of different optimization techniques on falsification problems. The evaluation shows that there is a significant difference in the efficiency of the falsification process. SNOBFIT~\cite{huyer2008snobfit}, a model-based approach, demonstrated the best performance for falsification in~\cite{Eddeland2020-ARCH}. Still some benchmark problems were not possible to falsify using this approach. On the basis of the observation of SNOBFIT's performance, it was discovered that SNOBFIT often explores new parameters at the extremes of parameter ranges (known as corner points) when it does not receive any hints from the quantitative semantics as to which direction to continue its search. This motivated the paper~\cite{Ramezani2021_Line} in which a new falsification method, Line-Search falsification (\LineSearch)~\cite{Ramezani2021_Line}, is proposed. \LineSearch is a direct-search method specifically developed for falsification problems and has demonstrated better performance than SNOBFIT as described in ~\cite{Ramezani2021_Line}. An explanation of the good performance of \LineSearch is the ability of \LineSearch to merge random exploration with local search by creating random lines in the parameter space and optimizing over line segments as well as the ability to investigate corner points while searching. Although \LineSearch outperforms previous approaches in falsifying benchmark problems, it is a direct-search method that does not attempt to learn a surrogate model of the objective function. \LineSearch will, for this reason, be compared with \BO methods since it has exhibited strong performance in solving falsification problems, as discussed in~\cite{Ramezani2021_Line}.

\BO has been previously explored for falsification of \CPSs in~\cite{Deshmukh2017Bayesian,Mathesen2021,Akazaki2016falsification,Silvetti2017active}. In~\cite{Mathesen2021}, \BO was adapted for use with conjunctive requirements, focusing on solving problems with numerous requirements. In~\cite{Akazaki2016falsification}, the Gaussian process upper confidence bound (GP-UCB) was employed for falsifying conditional safety properties, which require a safety property to hold whenever an antecedent condition is met. Gaussian process regression was utilized to identify the input search space region where the antecedent condition holds. The GP-UCB algorithm for conditional safety was further enhanced in~\cite{Silvetti2017active} by concentrating on points satisfying the antecedent. The GP-UCB approach is suitable for moderate-dimensional spaces, up to approximately 10 dimensions. However, for high-dimensional parameter spaces, this method is not feasible.

ALEBO~\cite{alebo} addresses REMBO's limitations, such as nonlinear distortion in the objective function and a low probability of containing an optimum within a linear embedding, by using a Mahalanobis kernel and sampling a linear embedding from the unit hypersphere. HeSBO~\cite{nayebi2019} offers an alternative by defining a linear embedding through hashing and sketching. Nevertheless, the performance of both methods still heavily depends on the optimal selection of the linear embedding size, which is difficult to determine in practice.

% Bayesian Optimization
\section{Background}
\label{Sec:Backgrould-Falsification-BO}
% Falsification of Cyber-Physical Systems
\subsection{Falsification of Cyber-Physical Systems}

%Finding a falsified point in black-box functions that, for example, describe a CPS is often a challenging task that requires many expensive simulations.

\begin{figure} 
    \centering
    \includegraphics[width=\textwidth]{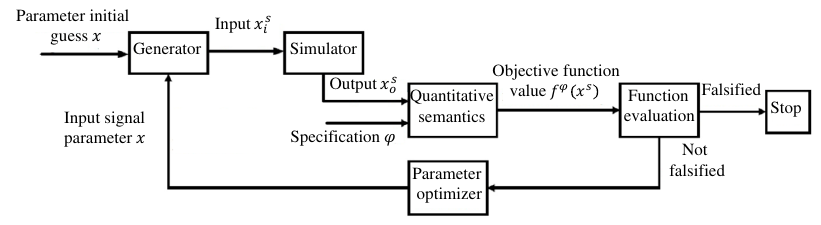}
    \caption{The process of optimization-based falsification~\cite{Claessen2018}.
    }
    \label{fig:falsification}
\end{figure}

The process of optimization-based falsification is shown in Figure~\ref{fig:falsification}. Initially, a generator creates input signals to the system based on an input parametrization. The input parameters $x\in\mathbb{R}^n$ with $n$, as the number of parameters, are used to generate an input trace describing a sequence of input vectors, $\xtracein[k]$, where $k$ ranges from the start to the end of the simulation, for the input ($i$) signals $(s)$. For example, a sinusoidal signal can be parameterized using the amplitude and the period or a piecewise constant signal by the values and time instants at which the signal changes value. Next, a simulator generates simulation traces of output signals $\ytraceout$, where the \SUT is simulated with the $\xtracein$ as inputs. The combination of both $\xtracein$ and $\ytraceout$, i.e. $\xtrace$, are used with the specification $\varphi$, possibly containing temporal operators, to evaluate the specification using a quantitative semantics.

% Quantitative semantics
A quantitative semantics determines whether the specification is satisfied and a measure of to what extent the specification is fulfilled. 
If the specification is falsified, the process ends. However, if it is not falsified, a parameter optimizer generates a new set of parameters for the input generator, and a new simulation of the system takes place.
Mapping between the input parametrization and the objective value $y\in\mathbb{R}$ of the selected quantitative semantics is taken as a black-box function $f^\varphi(x^s):\mathcal{X}\longrightarrow \mathbb{R}$. In falsification, $y=f^\varphi(x^s)<0$ denotes that the specification is falsified. Thus, minimizing the objective function $f^\varphi(x^s)$ should guide us to a point with $f^\varphi(x^s_{\rm falsified})<0$ if the system is possible to falsify. With slight abuse of notation, we will write $f(x)$ as shorthand for $f^\varphi$. 

% if something should be said here, it is the introduction of arg min that is not done previously.
%Thus the aim in falsification is to find to verify if there exists a $x \in \mathcal{X}$, such that $f(x) < 0$.
%the global minimum of the objective function within the domain $\mathcal{X}$. 
%Note that the $\arg\min$ of a black-box function returns a set since more than one value might achieve the minimum.

In falsification, the specification is often modeled using signal temporal logic (STL) or metric interval temporal logic (MITL) with their quantitative semantics defined in~\cite{Donze2010b} and~\cite{Fainekos2009}, respectively. In this paper, STL specifications are used.

\subsection{Signal Temporal Logic}

The syntax of STL~\cite{Maler2004} is defined as follows:
\noindent
\begin{equation*}
\varphi :: = \mu \,|\, \neg\mu \,|\,  \varphi \wedge \psi \,|\,  \varphi \vee \psi \,|\, \square_{[a,b]} \psi \,|\, \lozenge _{[a,b]} \psi \, |\, \varphi\,\mathcal{U}_{[a,b]} \psi,
\end{equation*}
where the predicate  $\mu$ is  $\mu  \equiv \mu (\xtrace)>0 $ and $\xtrace$ is a signal; $\varphi$ and $\psi$ are STL formulas; $\square_{[a,b]}$ denotes the \textit{globally} (\textit{always}) operator between times $a$ and $b$ (with $a<b$); $ \lozenge _{[a,b]}$ denotes the \textit{eventually} operator between $a$ and $b$; and $\mathcal{U}_{[a,b]}$ denotes the \textit{until} operator between $a$ and $b$.

The satisfaction of the formula $\varphi$ for the discrete signal $\xtrace$, consisting of both inputs and outputs to the \SUT, at the discrete-time instant $k$ is defined: %as: 
\noindent
\begin{align*}
    &(\xtrace,k) \models \mu &\Leftrightarrow \ \ &\mu(\xtrace[k]) > 0\\
    &(\xtrace,k) \models \neg\mu &\Leftrightarrow \ \ &\neg((\xtrace,k) \models \mu)\\
    &(\xtrace,k) \models \varphi \land \psi &\Leftrightarrow \ \ &(\xtrace,k) \models \varphi \land (\xtrace,k) \models \psi\\
    &(\xtrace,k) \models \varphi \lor \psi &\Leftrightarrow \ \ &(\xtrace,k) \models \varphi \lor (\xtrace,k) \models \psi\\
    &(\xtrace,k) \models \square_{[a,b]}\varphi &\Leftrightarrow \ \ &\forall k' \in [k+a,k+b], (\xtrace,k') \models \varphi\\
    &(\xtrace,k) \models \lozenge_{[a,b]}\varphi &\Leftrightarrow \ \ &\exists k' \in [k+a,k+b], (\xtrace,k') \models \varphi\\
    &(\xtrace,k) \models \varphi \ \mathcal{U}_{[a,b]}\psi &\Leftrightarrow \ \ &\exists k' \in [k+a,k+b] \ \ (\xtrace,k') \models \psi\\ \nonumber
    & & &\land \forall k'' \in [k,k'), (\xtrace,k'') \models \varphi \\
\end{align*}
Two quantitative semantics defined for STL are \maxS and \additiveS. Both of these can be expressed in terms of 
VBools~\cite{Claessen2018}, but in this paper, we focus on using \maxS since it is the most widely used quantitative semantics in falsification. A VBool $\langle v,z \rangle$ is a combination of a Boolean value $v$ (true $\top$, or false $\bot$) together with a real number $z$ that is a measure of how true or false the VBool is. 
This real value will estimate how convincingly a test passed or how severely it failed. The quantitative semantics defines this value. %In this paper, the \maxS semantics is used. 

% Quantitative Semantics:
\subsection{Quantitative Semantics}
\label{Subsec: Quantative_Semantics}
%Using VBools,  is defined, which is essentially the same as the standard STL quantitative semantics. 
For \maxS semantics, the \emph{and} ($\land$), \emph{or} ($\lor$), \emph{always} ($\square$), \emph{eventually} ($\lozenge$), and \emph{until} ($\mathcal{U}$) operators are introduced.

The \emph{and}-operator is defined as
\begin{align*}
\label{eq:and-additive}
   (\top, s) \land (\top, z) &= \Big(\top, \min(s,z) \Big), \hfill \notag \\
    (\top, s) \land (\bot, z) &= (\bot, z),  \\ \notag
    (\bot, s) \land (\top, z) &= (\bot, s), \\ \notag
    (\bot, s) \land (\bot, z) &= \Big(\bot, \max(s,z)\Big). \notag
\end{align*}

Using the de~Morgan laws, the \emph{or} operator can be defined in terms of \emph{and}, as: $(v_s,s)\lor (v_z,z) = \lnotvb (\lnotvb(v_{s}, s)  \land \lnotvb(v_{z}, z))$,
where VBool negation is defined as 
$\lnotvb (v_s,s) = (\neg v_s,s)$.

The \emph{always} operator over an interval $\left[a,b\right]$ is straightforwardly defined in terms of \emph{and}-operator, as 
\begin{align*}
  \square_{\left[a,b\right]}\varphi = \bigwedge\limits^b_{k=a}\varphi\left[k\right],
\end{align*}
\noindent
where $\varphi$ is  a finite sequence of \vbools defined for all the discrete-time instants in $\left[a, b\right]$. 

Furthermore, the \emph{eventually} operator is for both semantics defined over an interval $\left[a,b\right]$ in terms of \emph{always}-operator, as: $\lozenge_{\left[a,b\right]}\varphi = \neg(\square_{\left[a,b\right]}(\lnotvb\varphi)).$

% Max semantics- Until
Finally, the \emph{Max until}-operator as:
\begin{align*}
    \varphi \ &\mathcal{U}_{[a,b]} \ \psi
    = \sideset{}{}\bigvee \limits_{k=a}^{b}\left(\psi[k] \land \left(\sideset{}{}\bigwedge\limits_{k'=0}^{a-1}\varphi[k']\right)\right).
\end{align*}

% BO
\subsection{Bayesian Optimization}

\BO~\cite{Shahriari2016} assumes a probabilistic belief about the mapping between the input parametrization $x$ and the objective value $y$. Designing an acquisition function guides the optimization procedure to the optimal solution by selecting adequate configurations for evaluation.
While different regression models such as random forest can work as a surrogate model, a probabilistic model in \BO is typically based on Gaussian processes regression~\cite{rasmussen}. In principle, Gaussian process regression defines the output of a simulation-based falsification as $p(y) = \mathcal{GP}(y;\mu^\prime,K)$,
where $\mu^\prime$ is a prediction mean value, and $K$ is a covariance that returns similarity between points where $\sigma(x)=\sqrt{K(x,x)}$ is the marginal standard deviation of $f(x)$.  
%Each configuration $x\in\mathbb{R}^n$ in the probabilistic model comes with the prediction mean value $\mu(x)$ and $\sigma(x)=\sqrt{K(x,x)}$ which is the marginal standard deviation of $f(x)$. $\sigma(x)$ refers to a measure of uncertainty, also called a kernel. 

%approximation of $f(x)$, and $K$ is the covariance. 
%Given $n$ number of observations $\mathcal{D} = (x_i, y_i)_{i=1,\cdots,n}$, we can condition the distribution $p$ as $p(y|\mathcal{D}) = \mathcal{GP}(y;\mu_{y|\mathcal{D}},K_{y|\mathcal{D}})$.

%\begin{equation}\label{gp:eq}
%    p(y|\mathcal{D}) = \mathcal{GP}(y;\mu_{y|\mathcal{D}},K_{y|\mathcal{D}}).
%\end{equation}

% BO Algorithm:
\begin{algorithm}[ht] \small
%\captionsetup{labelformat=default}
\setcounter{algorithm}{0}
\caption{\bf Bayesian Optimization for falsification}
\label{Alg:BO}
\begin{algorithmic} [1]

\STATE \textbf{Input:} Input space $\mathcal{X}\subseteq\mathbb{R}^n$, the initial design size $M$, the max number of simulations $N$. \label{Alg:Input}
\STATE \textbf{Output:} A falsified design - if possible, to find within the given simulation budget.  \label{Alg:Output}
% Initial parameters and values:
\STATE $\{x_i\}^M_{i=1}\sim \mathbb{U}(x), \{y_i \longleftarrow f(x_i)\}^M_{i=1}$, \COMMENT{\textit{Sample random configurations from the uniform distribution and evaluate using simulation.}}  \label{Alg:Initialize}

\IF{ the specification is falsified at one of $\{x_i\}^M_{i=1}$} \label{Alg:IfFalsInitSample}
\RETURN The falsified point \COMMENT{\textit{Falsification is successful.}}  \label{Alg:IfFalsInitSampleFalsified}
\ELSE \label{Alg:IfFalsInitSampleElse}

\STATE $\mathcal{D}_0\longleftarrow\{(x_i,y_i)\}^M_{i=1}$ \COMMENT{\textit{Collect the initial design set.}}  \label{Alg:InitSet}
% While loop:
\FOR{(j = 1, 2, \dots, N)} \label{Alg:For}

\STATE $x^* \longleftarrow  \operatorname*{argmin}_{x\in\mathcal{X}} \, \alpha (x, \mathcal{D}_{j-1})$ \COMMENT{\textit{Train a probabilistic model as $p(y|\mathcal{D}) = \mathcal{GP}(y;{\mu^\prime}_{y|\mathcal{D}},K_{y|\mathcal{D}})$  \label{Alg:TrainModel}
with $\mathcal{D}_{j-1}$ and find the next configuration $x^*$} by minimization of the acquisition function $\alpha$.}
\STATE $y^*\longleftarrow f(x^*)$ \COMMENT{\textit{Evaluate the objective function by using simulation followed by the quantitative semantics.}} \label{Alg:EvaluateObj} 
\IF {$y^*<0$}   \label{Alg:IfFalsified}
  \RETURN $x^*$ \COMMENT{\textit{Falsification is successful.}}  \label{Alg:FalsificationOutput}
\ELSE  \label{Alg:ElseNotFalsified}
    \STATE $\mathcal{D}_{j}=\mathcal{D}_{j-1} \cup \{(x^*,y^*)\}$ \COMMENT{\textit{Update the sample set.}}  \label{Alg:UpdateSampleSet}
  \ENDIF  \label{Alg:EndIfFalsified}
\ENDFOR  \label{Alg:EndFor}
\ENDIF
\end{algorithmic}
\end{algorithm}

% BO presented in Alg.1
We summarize the main steps of \BO for falsification in Algorithm~\ref{Alg:BO}. At the start, we need to evaluate the objective function $f(x)$ on an initial number of samples (points) that is noted here as $M$ numbers, line~\ref{Alg:Input}. Defining the initial points in the search space $\mathcal{X}\subseteq\mathbb{R}^n$, where $n$ is the dimensional parameter space, is typically done by randomly selecting parameters within the allowed range, line~\ref{Alg:Initialize}. After evaluating $f(x)$, if the specification is falsified at any of the initial samples, lines~\ref{Alg:IfFalsInitSample}-\ref{Alg:IfFalsInitSampleFalsified}, the algorithm terminates with the falsified point. Otherwise, we create the initial sample set $\mathcal{D}_0\longleftarrow\{(x_i,y_i)\}^M_{i=1}$ based on an initial probabilistic model, line~\ref{Alg:InitSet}. 
%is defined using Eq.~\eqref{gp:eq}. 
Using the trained probabilistic model, we can decide which configuration to evaluate next. This process is done by minimization of the used acquisition function which will be introduced in~\ref{Turbo}.

% Acquisition Function
A key challenge is a trade-off between exploring regions of the parameter space not yet explored versus investigating areas around the most promising parameter values yet found. This trade-off is called the exploration-exploitation dilemma. Balancing adequately between exploring yet unexplored areas and exploiting promising regions determines the efficiency of the falsification process. Too much exploitation results in a greedy optimization where a surrogate model can easily be trapped in a local minimum. Vice versa, too much exploration would result in an inefficient performance where a surrogate model evolves with every new iteration without any exploitation. Therefore, an acquisition function that balances exploration vs. exploitation is used to select the next configuration for evaluation.
%It is an optimization problem itself.
Since an acquisition function is built on top of a surrogate model with a clear analytical form, optimizing it can be done efficiently. In general, acquisition functions come with different concepts in exploring and exploiting, which we discuss in Section.~\ref{Sec: Method}. 

%A key challenge is a trade-off between exploring regions of the parameter space not yet explored versus investigating areas around the most promising parameter values yet found. This trade-off is called the exploration-exploitation dilemma. 
%Each configuration $x\in\mathbb{R}^n$ in the probabilistic model comes with the prediction mean value $\mu(x)$ and $\sigma(x)=\sqrt{K(x,x)}$ which is the marginal standard deviation of $f(x)$. $\sigma(x)$ refers to a measure of uncertainty, also called a kernel. 

%Hence, 
The algorithm selects the next configuration, point $x^*$, for evaluation by optimizing a predefined acquisition function, line~\ref{Alg:TrainModel}. Once evaluated by simulating the \SUT with the generated input from the parameters $x^*$, the corresponding objective value $y^*$ is checked, line~\ref{Alg:EvaluateObj}. If $y^*<0$, we have found a falsifying configuration, and the optimization procedure ends, lines~\ref{Alg:IfFalsified}-\ref{Alg:FalsificationOutput}. Otherwise, the sample set $\mathcal{D}$ is updated, lines~\ref{Alg:ElseNotFalsified}-\ref{Alg:UpdateSampleSet}, and the previous steps are repeated. This continues until the total number of simulations, $N$, is exhausted, lines~\ref{Alg:For}-\ref{Alg:EndFor}.

%by exploring the search space of the input parametrization and/or reducing the uncertainty of a surrogate model (i.e., exploitation). 

% Acquisition Function
%Once a surrogate model is built, the objective is to find a configuration that improves our search. Balancing adequately between exploring yet unexplored areas and exploiting promising regions determines the efficiency of the falsification process. Too much exploitation results in a greedy optimization where a surrogate model can easily be trapped in a local minimum. Vice versa, too much exploration would result in an inefficient performance where a surrogate model evolves with every new iteration without any exploitation. Therefore, an acquisition function that balances exploration vs. exploitation is used to select the best-next configuration for evaluation.
%It is an optimization problem itself.
%Since an acquisition function is built on top of a surrogate model with a clear analytical form, optimizing it can be done efficiently. In general, acquisition functions come with different concepts in exploring and exploiting, which we discuss below. 
%Depending on an application, one acquisition function can be favored over another. 

% Method
\section{Method}
\label{Sec: Method}
% High-dimensional
The cost of using the Gaussian process for the surrogate model scales cubic with the number of samples evaluated, see~\cite{rasmussen}. High-dimensional applications typically require more evaluations to converge, so using vanilla BO in this setting becomes impractical~\cite{nayebi2019,frazier2018}. Furthermore, it is noted that BO searches more the edges, the lower and upper bounds of the input ranges, of the search space in high-dimensional applications, this results in suboptimal performance~\cite{Wang2016REMBO,Eriksson2019turbo}. In~\cite{Eriksson2019turbo}, this is mitigated by building multiple local probabilistic models in an approach called \turbo. 

%Therefore, an idea is to utilize a local search \BO method to exploit the objective function locally as done in~\cite{Eriksson2019turbo}. 

% turbo
\subsection{TuRBO}
\label{Turbo}
\turbo, a trust-region (TR) BO method, utilizes a sequence of local optimization runs using independent probabilistic models to overcome the problem of overexploiting. Furthermore, an implicit multi-armed bandit strategy at each iteration addresses the global optimization where a local run is selected for additional evaluations. It is possible to represent a \TR as a sphere, a polytope, or a hyperrectangle, with its center located at the point of the lowest objective function found so far in the optimization process. \turbo uses  Gaussian process (GP) models within a hyperrectangle \TR. A hyperrectangle centered at the current best solution is created with the predefined length. Within the hyperrectangle, a local surrogate model is trained. 
Large enough \TR would be equivalent to standard global \BO methods. Therefore, \TR should be large enough to encompass good solutions while remaining small enough to build an accurate local model. Consequently, there are limitations for the size of \TR $(L_{min}, L_{max})$. The \TR is expanded when a new point with a better objective function is found in that region; otherwise, \turbo shrinks when it appears stuck.
At the beginning of the \turbo process, a base side length is initialized for \TR, $L_{init}$. An acquisition function is used at each iteration, $i$, to select a batch of $q$ candidates $x_{1}^{(i)}, \dots, x_{q}^{(i)}$ within \TR. If better points are searched consecutively within \TR, the size of \TR is doubled, i.e., $\min(L_{min}, 2L)$. If \turbo fails to find better points, \TR is halved in size, $L/2$. If the size of \TR is less than $L_{min}$ or greater than $L_{max}$, the current \TR is discarded, and a new \TR with $L_{init}$ is initialized. The evaluated \turbo method in this paper uses a single local BO strategy using a \TR method in each search. \turbo uses the best current point with the lowest objective function that has been found so far as the most promising within a local optimization run instead of just doing random restarts. This leads to a more efficient use of the evaluation budget.

Using standard acquisition functions, \turbo finds the best-next configuration $x^*$ for evaluation. If the best-next evaluation $f(x^*)$ is better than the current best solution, then the trust region is expanded. Otherwise, it is shrunk. The default acquisition function used in \turbo is Thompson sampling (\TS).
To demonstrate the usage of \turbo in falsification and preferably w.r.t. the global \BO methods, our work covers a detailed study on the selection of an acquisition function. As part of this work, \turbo has been modified to work with Lower Confidence Bound (LCB)~\cite{Srinivas2010UCB}, and a version of Probability of improvement (\PI)~\cite{Kushner1964PI}. %, which will be introduced below.

% Acquisition Functions
% Thompson Sampling
\subsubsection{Thompson Sampling (\TS)}

\TS~\cite{Thompson1933likelihood} is a simple yet effective approach for handling the exploration-exploita\-tion dilemma in Bayesian optimization. Once we have a trained surrogate model, the concept is to greedily sample a configuration from the posterior with the lowest value, and sampling from the posterior generates TS's randomness.

% Lower confidence bound
\subsubsection{Lower Confidence Bound (LCB)}
Each prediction made by a surrogate model comes with the confidence interval explained with a corresponding standard deviation. The standard deviation is a measure of uncertainty. Thus, \LCB has been introduced to leverage this measure for exploration and exploitation. It refers to the lower bound of the uncertainty of the surrogate model. In this paper, we consider the minimization problem where the lower bound is of interest; for maximization problems, the Upper Confidence Bound (UCB) is used instead. The best-next configuration $x^*$ is~\cite{Srinivas2010UCB}
%finding minimizing~\cite{Srinivas2010UCB}

\begin{equation}
    \alpha_{\rm LCB}(x^*) \in\operatorname*{argmin}_{x\in\mathcal{X}}\mu^\prime(x) - \beta \cdot\sigma(x),
\end{equation}
\noindent
where the parameter $\beta \in \mathbb{R} \geq 0 $ balances exploitation and exploration. Note that the $\arg\min$ of a black-box function returns a set since more than one value might achieve the minimum. 
Small $\beta$ values mean more greedy 
exploitation, while large values mean more exploration. % and improving a surrogate model. 
Defining an optimal value for $\beta$ is an open question, in this work, we use a well-known formulation used in practice~\cite{Nardi2019practical}, $\beta = \sqrt{0.125\log(2 j + 1)}$, where $j$ is the number of simulations. 

%\begin{equation}
%    \beta = \sqrt{0.125\log(2 N + 1)},
%\end{equation}

% PI
\subsubsection{Probability of Improvement (PI)}
\PI~\cite{Kushner1964PI} defines the probability that the best-next configuration $x^*$ leads to an improvement with respect to a target value $\tau$, which ideally can be seen as the optimal solution $f_{\rm min}$. Thus, we write
\begin{equation}\label{pifun}
    \mathbf{P}(f(x)<\tau) = \mathbf{P}\Bigg(\frac{f(x) - \textcolor{black}{\mu^\prime}(x)}{\sigma(x)} < \frac{\tau - \textcolor{black}{\mu^\prime}(x)}{\sigma(x)}\Bigg) = \Phi\Bigg(\frac{\tau - \textcolor{black}{\mu^\prime}(x)}{\sigma(x)}\Bigg),
\end{equation}
where $\Phi$ is the standard normal cumulative distribution function. As the optimal solution $f_{\rm min}$ is unknown, $\tau$ is typically defined as the current best solution, although for falsification a negative value is required. Thus, if a potential configuration has an associated predicted value larger than the current best solution, then the optimization procedure is not improving. Scaling the difference between the current best solution and a prediction value with the standard deviation creates the exploitation nature of \PI. Because of maximizing the improvement, the next-best $x^*$ is
%Since we want to maximize the improvement, the best-next configuration $x^*$ is found
\begin{equation}\label{Eq:PI}
    \alpha_{\rm PI}(x^*) \in\operatorname*{argmax}_{x\in\mathcal{X}}\,\Phi\Bigg(\frac{\tau - \textcolor{black}{\mu^\prime} (x)}{\sigma (x)}\Bigg).
\end{equation}
The fraction in~\eqref{Eq:PI} is sometimes referred to as the \emph{U}-function~\cite{Echard2011ak}. Instead of maximizing~\eqref{Eq:PI}, we can minimize the \emph{U}-function as
\begin{equation}\label{Eq:U}
    \alpha_{\rm U}(x^*) \in\operatorname*{argmin}_{x\in\mathcal{X}} - \frac{\tau - \textcolor{black}{\mu^\prime} (x)}{\sigma (x)} = \operatorname*{argmin}_{x\in\mathcal{X}} \frac{\textcolor{black}{\mu^\prime}(x) - \tau}{\sigma (x)}.
\end{equation}
The formulation in~\eqref{Eq:U} with the absolute value of $|\textcolor{black}{\mu^\prime}(x^*)-\tau|$ is also found in reliability analysis, where the objective is to approximate the probability of failure. Using the absolute value here improves the threshold between failure and non-failure events. In reliability analysis, the best-next configuration $x^*$ close to $\tau$ from any side is sufficient for evaluating~\cite{Echard2011ak,Schobi2017rare}. As we are only interested in falsification for CPSs (i.e., having values less than 0), defining $\tau\leq0$ emphasizes failure events~\cite{Echard2011ak,Schobi2017rare}. %In terms of efficiency, selecting $\tau\ll0$ for falsified events far from 0 can result in even faster falsification. 
In the present study, we experiment with two choices, $\tau=0$ and $\tau=-1$. 

% Prior BO
%subsection{User Priors for the Optimum}
\subsection{Incorporating prior belief}
In certain situations, the practitioner has an available prior belief about the potential location of the optimum~\cite{Souza2021Prior,hvarfner2022pi}. While this source of information might be available, vanilla \BO fails to incorporate it. Therefore, the work previously done in~\cite{Souza2021Prior,hvarfner2022pi} proposed how to modify \BO to inject this prior. In particular, the latest algorithm $\pi$\BO~\cite{hvarfner2022pi} is conceptually simpler than the previous work. The objective is to modify an acquisition function by multiplying it with a predefined probability distribution $\pi(x)$ as

\begin{equation}\label{pibo}
    x^* \in \operatorname*{argmax}_{x\in\mathcal{X}} \alpha(x, \mathcal{D}_{j-1})\cdot\pi(x)^{\beta^*/j}.
\end{equation}
\noindent
for the $j$-th iteration with $j \in \{1, \ldots ,N \}$. $\beta^* \in \mathbb{R}^+$ is a hyperparameter that is used as the practitioner's confidence about the prior knowledge. 
Here, a probability distribution $\pi(x)$ serves to describe our belief about the optimum. For example, in falsification, falsified points for several benchmark problems are located at the edges of the search space, making it easy to falsify if this information is known~\cite{Ramezani2021_Line}. Therefore, in the present study, we defined $\pi(x)$ as a U-shaped distribution where the edges are weighted more than the inside area. Even though the prior can be wrong, \BO can still converge the optimal solution as a result of the forgetting factor in~\eqref{pibo} as proven in~\cite{hvarfner2022pi}. By raising $\pi(x)$ in~\eqref{pibo} to a power of $\beta^*/j$, the wrong prior decays towards zero with growing $j$. %Here, the hyperparameter $\beta^*$ is used as the practitioner's confidence about the prior knowledge.

% Evaluation
\section{Experimental Evaluation}
\label{Sec:experimental}

% ARCH Benchmark Examples
%\subsection{ARCH Benchmark Examples}
We evaluate the proposed \BO methods on benchmark problems from~\cite{Ernst2019-ARCH19:}, and~\cite{Ramezani2020-wodes} which are introduced briefly in appendix~\ref{Sec:Benchmark_Examples}.   %In the following, we refer to all these examples as ARCH benchmark examples. 
%These examples are introduced briefly in appendix~\ref{Sec:Benchmark_Examples}.  
For the benchmark problems in~\cite{Ernst2019-ARCH19:}, two variants of input signals are considered Instance 1 and Instance 2, respectively. Instance 1 allows arbitrary piecewise continuous input signals, but with a finite number of
discontinuities in the ARCH19 competition. On the other hand, Instance 2 is restricted to constrained input signals, i.e., the input signal format is fixed, but discontinuities are allowed. The problems such as \textit{AT}, \textit{CC}, \textit{NN}, and \textit{SC} include both Instance 1 and Instance 2 type, while the problem \textit{AFC}, \textit{WT}, and \textit{F16} do not include different types of instances.
Additionally, three benchmark problems from~\cite{Ramezani2020-wodes} are included, i.e., $AT^\prime$, modulator $\Delta - \Sigma$, and \textit{SS}. 
%Note that the example Automatic Transmission ($AT^\prime$) has different specifications and input ranges from the Automatic Transmission (\textit{AT}).

%\vspace{-5mm}
% Experimental Setup
\subsection{Experimental Setup}

The performance of the proposed \BO methods compared to vanilla \BO is compared against state-of-the-art methods found in the literature, such as \hybrid (i.e., an optimization-free method)~\cite{Ramezani2021_Line} and line-search falsification \LineSearch method (i.e., a direct-search optimization method). 
% Initial setups for each method
We set up the experiments using Breach~\cite{Donze2010}, with \maxS semantics~\cite{Donze2010b,Claessen2018}. As \BO methods require a set of initial samples to start the process, we set the initial number of samples to $2\cdot n$, where $n$ is the number of input parameters (i.e., the dimensionality of the optimization problem). This method is implemented in Python. In this implementation, the optimization process is conducted in Python, while the simulation and evaluation of the objective function are performed in MATLAB.
For $\pi$BO, a U-shaped distribution is used, which assigns a probability density to different points in the space: "prior:[0.4,0.1,0.1,0.4]". Similar to \turbo, $\pi$BO is implemented in Python and interacts with MATLAB.
For the \LineSearch method, the maximum number of iterations to improve a single line is set to three. The presented results for \LineSearch use Option 4, which works with lines extending beyond the boundaries of input ranges and thus has a higher chance of resulting in corner values; see~\cite{Ramezani2021_Line} for more details. The \hybrid method starts with a corner point, and the next point is a uniform random (UR) point. It switches between the corners and UR points until the maximum number of simulations, $N = 1000$, is reached or a falsified point is found. The number of corners is limited and depends on the dimensionality of the \SUT since there are $2^n$ corners. If the maximum number of corners is reached, \hybrid continues using only random input points. Both \LineSearch and \hybrid are implemented in MATLAB.
 
%\vspace{-2mm}    
% Tables   
In Tables~\ref{tab:Easy_to_Falsify}-\ref{tab:Hard_to_Falsify}, we present and discuss the benchmark problems where the choice of the optimization method affects the falsification process.
The results for the benchmark problems that are easily falsified with a few simulations regardless of which used optimization method are shown in Appendix~\ref{Other_Examples_Appendix}. 
%Tables~\ref{tab:Instance1}-~\ref{tab:Without_Instance} in Section~\ref{Other_Examples_Appendix}. 
We also include the results for the specifications that are hard to falsify regardless of the optimization methods, %Table~\ref{tab:Not_Falsified} 
in Appendix~\ref{Unfalsifiable_Examples}.   

% the proposed \BO methods can mostly exceed the performance of the state-of-the-art methods as shown 
In these tables, the first column denotes the specifications; the second refers to which instance is used to evaluate the specification. 
%(i.e., to which instance does belong). 
The third column shows the number of dimensions, and the rest of the columns contain the evaluation results for \VanillaBO, \turbo with %the different acquisition functions
\TS, \LCB, and \PI; $\pi$BO, \LineSearch, and \hybrid, respectively.
Two different target values are considered for \PI, \tauZero~, and \tauOne~. 
Each falsification is set to have 1000 maximum number of simulations. Since the falsification process contains random elements, we repeat the falsification process 20 times.
Two values are presented for each specification; the first is the relative success rate of falsification in percent. There are 20 falsification runs for each parameter value and specification; thus, the success rate is a multiple of 5\%. The second value, inside parentheses, is the average number of simulations (rounded) \emph{per successful falsification}.

% Easy Table: 
%\begin{landscape}
\begin{table*}[t] \scriptsize
\renewcommand{\arraystretch}{1}
\caption{Results for the problems that are easily falsified using an optimization-free method. Instances refer to different types of input parameterization and interpolation.
The first number is the relative success rate of falsification in percent, and the number in parenthesis is the average number of simulations (rounded) per successful falsification out of 1000 simulations.}
\label{tab:Easy_to_Falsify}
\centering
\begin{tabular}{c c c c c c c c c c c}

\toprule
Specifications
& Instances
& Number of 
& \multicolumn{1}{c}{vanilla} 
& \multicolumn{1}{c}{\turbo} 
& \multicolumn{1}{c}{\turbo} 
& \multicolumn{1}{c}{\turbo}
& \multicolumn{1}{c}{\turbo}
& \multicolumn{1}{c}{$\pi$BO} 
& \multicolumn{1}{c}{LSF} 
& \multicolumn{1}{c}{HCR} 
\\

&
& Dimensions
& \multicolumn{1}{c}{\BO} 
& \multicolumn{1}{c}{\TS} 
& \multicolumn{1}{c}{\LCB} 
& \multicolumn{1}{c}{\PI ($\tau = 0$)}
& \multicolumn{1}{c}{\PI ($\tau = -1$)}
& \multicolumn{1}{c}{} 
& \multicolumn{1}{c}{} 
& \multicolumn{1}{c}{} 
\\

\toprule

\multirow{10}{*}{}

$\varphi_6^{AT}$
&
Instance 1
& 
8
& 65 (322)
& 100 (90)
& 100 (84)
& 100 (143)
& 100 (68)
& \textbf{100 (55)} 
& 100 (63)
& 100 (117)
\\

$\varphi_7^{AT}$
&
Instance 1
&
8
& 100 (165)
& 95 (166)
& 100 (38)
& 100 (32)
& 100 (59)
& 100 (85)
& \textbf{100 (13)} 
& 100 (130)
\\

$\varphi_8^{AT}$
&
Instance 1
&
8
& 95 (143)
& 100 (32)
& 100 (79)
& 100 (64)
& 100 (70)
& 100 (242)
& \textbf{100 (26)}
& 100 (178)
\\

\midrule

$\varphi_2^{AT}$
&
Instance 2
&
40
& 100 (13)
& 95 (167)
& 100 (87)
& 100 (161)
& 100 (90)
& 100 (62)
& 100 (187) 
& \textbf{100 (3)}
\\
\midrule

$\varphi_3^{AT^\prime} (T = 4.5)$
&
-
&
10
& \textbf{100 (15)}
& 85 (300)
& 85 (312)
& 90 (289)
& 100 (176)
& 100 (82)
& 100 (173)
& 100 (21)
\\

$\varphi_4^{AT^\prime} (T = 1)$
&
-
&
10
& 100 (54)
& 65 (345)
& 100 (308)
& 70 (504)
& 95 (335)
& 75 (363)
& 60 (376)
& \textbf{100 (1)}
\\

$\varphi_5^{AT^\prime} (T = 1)$
&
-
&
10
& \textbf{100 (5)}
& 100 (301)
& 100 (219)
& 100 (174)
& 95 (323)
& 100 (29)
& 100 (37)
& 100 (120)
\\

$\varphi_8^{AT^\prime} (\bar{\omega} = 3500)$
&
-
&
10
& 100 (47)
& 35 (511)
& 30 (679)
& 25 (576)
& 20 (411)
& 100 (83)
& 55 (324) 
& \textbf{100 (5)}
\\

\midrule

$\varphi_2^{CC}$
&
Instance 2
&
40
& 100 (6)
& 100 (151)
& 90 (104)
& 75 (242)
& 100 (137)
& 100 (30)
& 90 (304)
& \textbf{100 (1)}
\\

\midrule

$\varphi_2^{NN}$
&
Instance 1
&
12
& 100 (283)
& 25 (445)
& 10 (415)
& 15 (502)
& 20 (345)
& 35 (232)
& 80 (253)
& \textbf{100 (83)}
\\

\midrule

$\varphi_1^{SS} (\gamma = 0.7)$ 
&
-
&
2
& 100 (18)
& 90 (393)
& 95 (405)
& 85 (235)
& 95 (435)
& 100 (18)
& 100 (66)
& \textbf{100 (3)}
\\

$\varphi_1^{SS} (\gamma = 0.8)$ 
&
-
&
2
& 100 (24)
& 60 (17)
& 65 (523)
& 70 (202)
& 50 (274)
& 100 (15)
& 100 (60)
& \textbf{100 (3)}
\\

$\varphi_1^{SS} (\gamma = 0.9)$
&
-
&
2
& \textbf{100 (52)}
& 35 (803)
& 55 (282)
& 0 (-)
& 30 (257)
& 100 (16)
& 100 (121)
& 100 (121)
\\

\bottomrule

\end{tabular}
\end{table*}

% Hard Table: 
\begin{table*}[t] \scriptsize
\renewcommand{\arraystretch}{0.8}
\caption{Results for the problems that are not easily falsified using an optimization-free method and require a large number of simulations for successful falsification. Instances refer to different types of input parameterization and interpolation.
The first number is the relative success rate of falsification in percent, and the number in parenthesis is the average number of simulations (rounded) per successful falsification out of 1000 simulations.}
\label{tab:Hard_to_Falsify}
\centering
\begin{tabular}{c c c c c c c c c c c}

\toprule
Specifications
& Instances
& Number of 
& \multicolumn{1}{c}{vanilla} 
& \multicolumn{1}{c}{\turbo} 
& \multicolumn{1}{c}{\turbo} 
& \multicolumn{1}{c}{\turbo}
& \multicolumn{1}{c}{\turbo}
& \multicolumn{1}{c}{$\pi$BO} 
& \multicolumn{1}{c}{LSF} 
& \multicolumn{1}{c}{HCR} 
\\

&
& Dimensions
& \multicolumn{1}{c}{\BO} 
& \multicolumn{1}{c}{\TS} 
& \multicolumn{1}{c}{\LCB} 
& \multicolumn{1}{c}{\PI ($\tau = 0$)}
& \multicolumn{1}{c}{\PI ($\tau = -1$)}
& \multicolumn{1}{c}{} 
& \multicolumn{1}{c}{} 
& \multicolumn{1}{c}{} 
\\

\toprule

\multirow{10}{*}{}

$\varphi_7^{AT}$
&
Instance 2
&
40
& 0 (-)
& 100 (200)
& 100 (242)
& 100 (185)
& \textbf{100 (184)}
& 5 (645)
& 100 (121)
& 0 (-)
\\

$\varphi_8^{AT}$
&
Instance 2
&
40
& 0 (-)
& 100 (287)
& 100 (300)
& 95 (359)
& 100 (334)
& 0 (-)
& \textbf{100 (127)}
& 0 (-)
\\

$\varphi_9^{AT}$
&
Instance 2
&
40
& 0 (-)
& 100 (248)
& 100 (225)
& 100 (215)
& 100 (195)
& 0 (-)
& \textbf{100 (75)}
& 0 (-)
\\

\midrule

$\varphi_6^{AT^\prime} (T = 10)$
&
-
&
10
& 0 (-)
& 95 (126)
& \textbf{100 (161)}
& 90 (191)
& 85 (196)
& 90 (293)
& 30 (543)
& 0 (-)
\\

$\varphi_6^{AT^\prime} (T = 12)$
&
-
&
10
& 15 (768)
& 100 (67)
& \textbf{100 (43)}
& 95 (78)
& 100 (64)
& 100 (228)
& 95 (173)
& 60 (297)
\\

$\varphi_7^{AT^\prime}$
&
-
&
10
& 0 (-)
& 45 (632)
& 25 (507)
& 45 (337)
& 35 (708)
& 35 (525)
& 25 (568)
& 30 (621)
\\

\midrule

$\varphi_4^{CC}$
&
Instance 1
&
8
& 0 (-)
& 75 (431)
& \textbf{80 (348)}
& 55 (365)
& 65 (351)
& 30 (629)
& 40 (381)
& 0 (-)
\\

$\varphi_4^{CC}$
&
Instance 2
&
40
& 0 (-)
& 45 (620)
& \textbf{95 (717)}
& 0 (-)
& 60 (808)
& 0 (-)
& 10 (620)
& 0 (-)
\\

\midrule

$\varphi_2^{NN}$
&
Instance 2
&
3
& 35 (230)
& 10 (343)
& 15 (570)
& 5 (135)
& 5 (191)
& 25 (299)
& \textbf{45 (400)}
& 0 (-)
\\

\midrule

$\varphi_1^{\Delta - \Sigma} U \in [−0.35, 0.35]$
&
-
& 
4
& \textbf{100 (108)}
& 95 (156)
& 90 (124)
& 95 (122)
& 95 (149)
& 100 (149)
& 100 (249)
& 0 (-)
\\

\midrule

$\varphi_1^{F16}$
&
-
& 
3
& 25 (685)
& 70 (515)
& 70 (303)
& \textbf{85 (337)}
& 80 (356)
& 60 (350)
& 65 (530)
& 5 (767)
\\

\bottomrule

\end{tabular}
\end{table*}
%\end{landscape}

% Results
\subsection{Results}
While Table~\ref{tab:Easy_to_Falsify} includes the benchmark problems that can be easily falsified using a straightforward approach such as \hybrid, Table~\ref {tab:Hard_to_Falsify} covers the hard problems and specifications where the number of simulations to falsify easily exceeds the maximum simulation budget~\cite{Ramezani2021_Line}.
% Easy benchmark examples
In particular, in Table~\ref{tab:Easy_to_Falsify}, these selected specification are $\varphi_6^{AT}$-$\varphi_8^{AT}$ for Instance 1 of $AT$ problem; Instance 2 of $\varphi_2^{AT}$; $\ATprime{3}(T = 4.5)$, $\ATprime{4}(T = 1)$, $\ATprime{5}(T = 1)$, $\varphi_8^{AT^\prime} (\bar{\omega} = 3500)$; Instance 2 of $\varphi_2^{CC}$; Instance 1 of $\varphi_2^{NN}$ and all specifications of \SS~problem. 

%As seen for \AT example, the proposed \BO methods (\turbo regardless of a selected acquisition function and $\pi$BO) with the state-of-art method \LineSearch perform better than \VanillaBO and other state-of-the-art methods. 
For \AT problem, \turbo (regardless of a selected acquisition function), $\pi$BO, and \LineSearch perform better than \VanillaBO.
Vanilla BO is not successful in falsifying $\varphi_6^{AT}$ and $\varphi_8^{AT}$ in each run.
However, in Instance 2 of \AT problem, \VanillaBO for $\varphi_2^{AT}$ of this problem performs as well as \hybrid and better than other optimization-based methods. %When \VanillaBO is used for high-dimensional applications, exploration of the edges is more frequent because optimizing an acquisition function is difficult and prone to significant errors~\textcolor{red}{\cite{Eriksson2019turbo}}.
When vanilla BO is used for high-dimensional applications, exploration of the edges is more frequent. Search spaces expand faster than sampling budgets resulting in regions with high posterior uncertainty which are typically located at the edges due to the extrapolation of a model. Hence, a typical acquisition function would overemphasize the edges and fail to exploit promising areas.
Hence, \VanillaBO searches for more edges and likely more corners. As a result, \VanillaBO falsifies $\varphi_2^{AT}$ within a few simulations. On the other hand, $\pi$BO also shows a good performance here because it emphasizes more the corners following our prior U-shaped distribution. \turbo is prone to explore fewer corner points as the search is limited to local trust regions. In particular, \TS is not always successful, while \PI and \LCB require more simulations than \hybrid and \VanillaBO. 

For the specifications of $AT^\prime$ problem, \VanillaBO and $\pi$BO, except for $\ATprime{4}(T = 1)$, also perform quite well with a 100\% success rate. Further, the \turbo method and \LineSearch are not
always successful for the benchmark problems $\ATprime{3}(T = 4.5)$, $\ATprime{4}(T = 1)$, $\ATprime{5}(T = 1)$, $\varphi_8^{AT^\prime} (\bar{\omega} = 3500)$. In contrast, \VanillaBO falsifies them with only a few simulations. 
Also, for $\varphi_5^{AT^\prime} (T = 1)$, \VanillaBO and
$\pi$BO beats \hybrid with fewer evaluations. While \VanillaBO required 5 simulations on average to falsify, $\pi$BO needed 29 simulations. The better performance of \VanillaBO and $\pi$BO results in $\varphi_8^{AT^\prime} (\bar{\omega} = 3500)$ performing among the top optimization-based methods, but not better than \hybrid that only requires 5 simulations.

% CC, \varphi_2^{CC}, Instance 2
For $\varphi_2^{CC}$ of instance 2, $\pi$BO and \VanillaBO work similarly to \hybrid and much better than other optimization-based methods. This specification can be falsified at some of the corner points. 
Among \turbo results, \PI (\tauOne~) and \TS perform better than \LCB and \PI (\tauZero~), which are successful only 90\% and 75\%, respectively.  

% NN, $\varphi_2^{NN}$, Instance 1
The benchmark problem $\varphi_2^{NN}$ is falsified with \hybrid with 100\% and using 83 simulations on average. In contrast, \VanillaBO is the only optimization-based method that is successful in each independent trial. Based on the evaluation of falsified points, we observe that this specification is falsifiable where at least some input parameters are at the upper or lower bound of input ranges. Hence, \VanillaBO and \hybrid performed better than \turbo. \turbo never evaluates the corner points, or the points are on the bound of the input ranges. However, $\pi$BO does not falsify efficiently, only 35\% are successfully falsified; it might be because the number of dimensions is increased, and hence the efficiency of $\pi$BO drops.  

% SS
\SS~is a two-dimensional synthetic problem that tricks the optimizing algorithms that try to estimate gradients to search in the wrong direction. For all specifications of \SS, \hybrid, \VanillaBO, \LineSearch, and importantly $\pi$BO provide better results than \turbo. 
Both input parameters are defined within the range $[-1, 1]$ in this special case. 
The target specification is falsified at the corner point $x^T=[1,1]$ and close to it. The gradient cannot point toward the falsification area if the initial samplings of \emph{First Input} and \emph{Second Input} are approximately in the ranges $(-1, 1)$ and $(-1, 0.8)$, respectively when $\gamma = 0.7$, which is a threshold parameter in this problem.
Hence, the three methods \hybrid, \VanillaBO, \LineSearch, and $\pi$BO that can search the corner points which lead that all falsify this problem better than \turbo. On the other hand, \turbo searches more within the input ranges. Thus, it is difficult for \turbo to approach the failure area with a local hyperrectangle trust region that is centered at the best solution found. A larger value for $\gamma$ such as $(\gamma = 0.8)$ and $(\gamma = 0.9)$ results in a smaller failure surface. Thus, the performance of \turbo drops additionally with the best performance with \PI (\tauZero~) as 70\% and \LCB as 55\% for $(\gamma = 0.8)$ and $(\gamma = 0.9)$, respectively. 
Next, we discuss the benchmark problems noted here as the hard problems because an optimization-free method such as \hybrid cannot falsify them. The results are provided in Table~\ref{tab:Hard_to_Falsify}.
% AT, specifications 7-9, Instance 2
For the specifications, 7-9, of \AT problem, \hybrid, \VanillaBO, and $\pi$BO cannot falsify them. Based on our evaluation of these specifications, they are not falsifiable close to corners or edges using the chosen input parameters and used optimization methods. In particular, emphasizing the corners as done in $\pi$BO is wrong. However, as a result of the forgetting factor, $\pi$BO still converges to the optimal solution as it was successful in only one run out of 20 runs in $\varphi_7^{AT}$. Like any other global \BO approach in a high-dimensional setting, the efficiency of $\pi$\BO eventually drops significantly as the bias of a Gaussian process to exploit the edges of the search space is emphasized. In \turbo, this shortcoming is handled, as shown in the results for the hard problems. %Because the training of a Gaussian process is done locally within a trust region that shrinks and expands based on the performance, the inherent presence of regions with large posterior uncertainty commonly found in the global \BO approach due to the curse of dimensionality is reduced to a minimum.
In \turbo, the training of a Gaussian process is done locally within a trust region that shrinks and expands based on the performance. Hence, the inherent presence of regions with large posterior uncertainty commonly found in the global \BO approach due to the curse of dimensionality is reduced.
The performance of \turbo is similar regardless of the acquisition function used. Compared to \LineSearch, more simulations are needed.

%AT', $\varphi_6^{AT^\prime}$
In the $AT^\prime$ problem, we can see a clear advantage of \turbo over other \BO methods and state-of-the-art approaches. 
For $\varphi_6^{AT^\prime}$, with both $(T = 10), (T = 12)$, \LCB is successful in each run, with 161 and 43 average on simulations respectively, while \LineSearch is only successful 30 and 95\%. 
Another \BO method, $\pi$BO, also demonstrates a good performance for $\varphi_6^{AT^\prime}$, with 95\% success rate when $(T = 10)$ and 100\% when $(T = 12)$. Because the dimensionality for these specifications is moderate, $\pi$BO forgets the wrong prior and eventually converges to the falsified areas. 
Vanilla BO is not successful in falsifying these specifications because simulations are mostly evaluated at the edges that are not falsifiable.

%\vspace{-1 mm}
%$\varphi_7^{AT^\prime}$
$\ATprime{7}$ is one specification that is a hard problem to be falsified with a specific feature.
The \STL formula of this specification is $\varphi_7^{AT^\prime} = \lnot\Big((\Box_{[0,1]}gear == 1) \land (\Box_{[2,4]}gear == 2) \land(\Box_{[5,7]}gear == 3) \land (\Box_{[8,10]}gear == 3) \land(\Box_{[12,15]}gear == 2)\Big)$.
In general, it is impossible to determine how close we are to fulfilling equality predicates, such as $gear==1$, $gear==2$, or $gear==3$. In other words, if the gear is 1, for example, the objective value using \maxS semantics will be a positive constant value if the specification is fulfilled, otherwise, a negative constant value. For simplicity, we assume that the positive constant value is 1 while the negative constant value is -1. 
In this specification, \emph{always}-operators are combined in one big conjunction. The first \emph{always}-operator checks to see whether $gear==1$ between $t=0~sec$ and $t=1~sec$ or not. 
If it is satisfied, the \emph{always}-operator has the total objective value $min([1, 1,\dots, 1]) = 1$, which means that the gear is always equal to 1 in $t = [0, 1]$. Similarly, for the $gear==2$ and $gear==3$ at other times, the specification has a positive constant value of $1$. On the other hand, if the specification is not satisfied at any time we will have something like $\min([1, -1,\dots, 1]) = -1$. Since the \maxS yields the same value for different inputs, model-based methods such as \BO cannot learn anything meaningful to sufficiently explore and exploit the given search space. The main assumption of \BO is that the target function should be sufficiently smooth.
\TS and \PI (\tauZero~)
provide slightly better performance compared to \LCB and \PI (\tauOne~) with a 45\% success rate. Because \TS puts more emphasis on exploitation than other methods, it is expected to show better performance. Other methods balance equally between exploitation and exploration, which is useless in this specific case.
Even though the learning is not adequate,% \BO methods expect \VanillaBO can still falsify 
\turbo, and $\pi$BO can still falsify this problem but not always, as shown in Table~\ref{tab:Hard_to_Falsify}.
%for a few independent trials.

% CC, $\varphi_4^{CC}$, both instances
For both instances in $\varphi_4^{CC}$, we see a significant advantage of using \turbo w.r.t. other optimization-based or optimization-free methods. For Instance 1, there is not much difference between the performance of \turbo with different acquisition functions, although \LCB performs slightly better with 80\%. \turbo performs much better than other optimization methods, e.g., 40\% increasing rate than \LineSearch.
For Instance 2, there is a difference between the performance of \PI when the target value is specified differently. By evaluating the objective value of the falsified point in each run of this specification, we could understand that the global optimum is close to and less than $-1$. Hence, having a target of $0$ does not provide good results. 
This specification was one of the hardest benchmark problems evaluated in~\cite{Ramezani2021_Line} where the proposed method, \LineSearch, was able to falsify $\varphi_4^{CC}$ in 10\% of the runs. As it can be seen from Table~\ref{tab:Hard_to_Falsify}, more simulations are needed to falsify this specification.
The acquisition function \LCB in 
\turbo needs fewer simulations for falsification than other methods and has a success rate of 95\%.
Based on the evaluation collected from this problem, we could conclude that the falsification, in this case, happens in a small region of the search space.  
Therefore, random approaches might not be a good choice to find this small region. A target of $-1$ is appropriate since the minimum achievable object value is close to $-1$. 

%could happen around this target rather than when the target is \tauZero~.

%% NN, $\varphi_2^{NN}$, Instance 2
In the benchmark problem $\varphi_2^{NN}$ (Instance 2), \LineSearch performs the best with the success rate of 45\%, exceeding especially \turbo with 5, 10, and 15\%. Based on the evaluation of the falsified point, there are reasons why all other methods are not successful in falsifying for each trial. First, this specification is falsifiable, where at least some input parameters are at the upper or lower bound of the input interval. The falsifiable region is small, and close to the boundaries.

%Moreover, the search space on these boundaries is too narrow. 

% Modulator
For the modulator problem, \VanillaBO and $\pi$BO slightly work better than others in falsifying each trial. These results demonstrate that $\pi$BO works well when the number of dimensions is relatively small. For $F16$, \turbo with \PI (\tauZero~) provides a better result with a success rate of 85\% compared to other methods.

\begin{figure*}[t]
	\centering
	\centerline{\includegraphics[height=8cm]{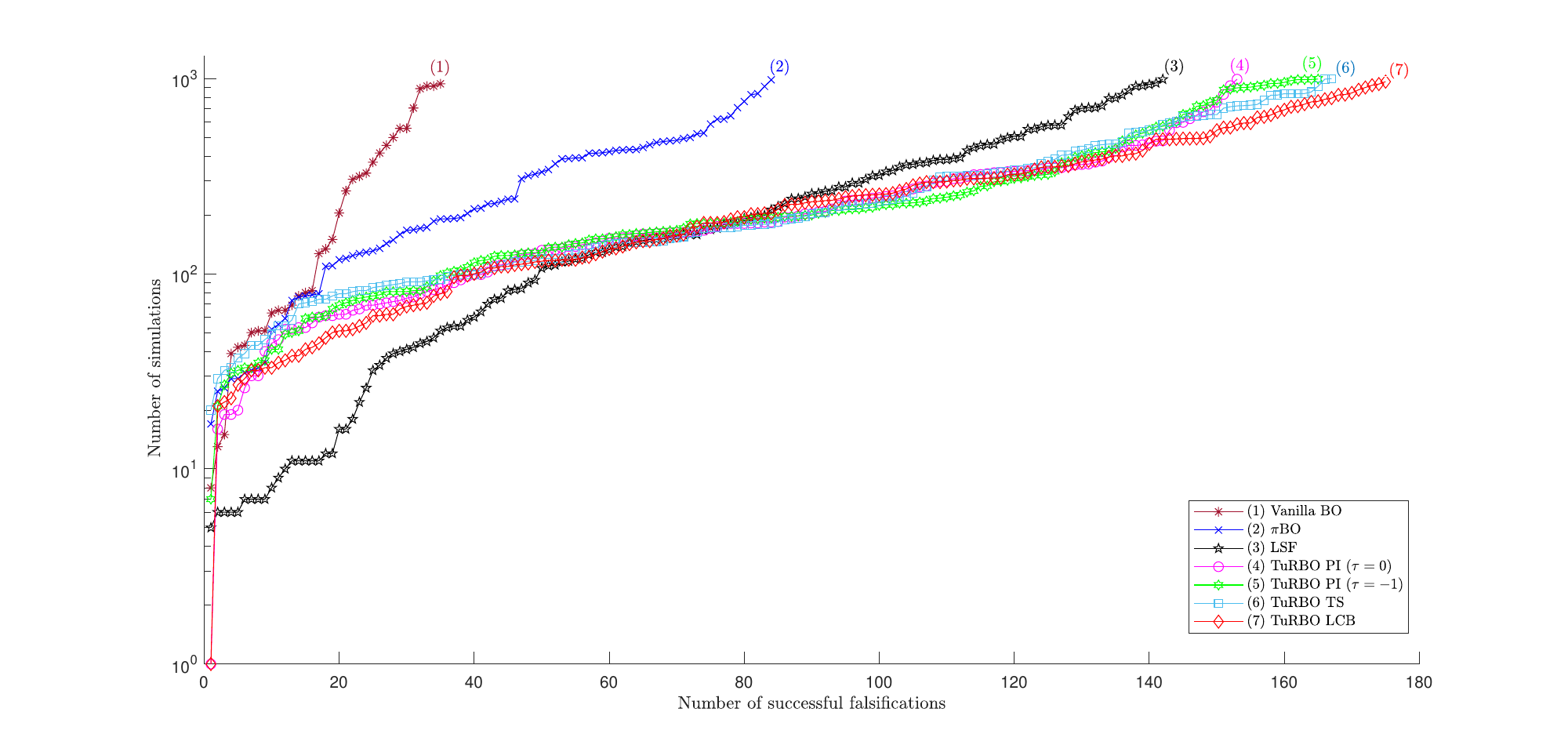}}
	\caption{A cactus plot showing the performance of optimization-based methods on the hard problems in Table~\ref{fig:cactus_Hard_Examples}. The plotted values show how many successful falsifications ($x$-axis) were completed in less than always simulations ($y$-axis, logarithmic scale).}
	\label{fig:cactus_Hard_Examples}
\end{figure*}

% Cactus plot for the hard examples
% To better comparison of the 
To examine and study the performance of different optimization-based methods, an aggregated comparison is provided in
Figure~\ref{fig:cactus_Hard_Examples} with the hard specifications presented in Table~\ref{fig:cactus_Hard_Examples}. 
As shown in Figure~\ref{fig:cactus_Hard_Examples} and evident from our early discussion, \turbo falsifies more benchmark problems with fewer simulations if we select \LCB w.r.t. other acquisition functions such as \PI (\tauOne~), \PI (\tauZero~) and the default \TS.
%\zahra{Lines140-141}
\turbo with \LCB falsified 5 specifications out of 11 that are specified as the hard problems in Table~\ref{tab:Hard_to_Falsify} with the success rate of 100\%, while \TS and \PI (\tauOne~) could falsify four specifications similarly as \LineSearch. \PI (\tauZero~) falsified 2 specifications similarly to $\pi$BO. On the other hand, \VanillaBO is successful only for one specification in each trial.    
By taking the average of these success rates, we can see that \LCB provides the best performance on average, with a success rate of 79.55\%. Other acquisition functions have the average success rate as follows \TS with 75.90\%; \PI (\tauOne~) with 75.45\%, \PI (\tauZero~) with 69.54\%, \LineSearch with 64.54\%, $\pi$BO with 38.18\%, and finally, \VanillaBO with 15.90\%.

% Discussion
\subsection{Discussion}

% Comparison among the evaluated methods
%As demonstrated in our experiments, the \BO method showed an effective and good performance for falsification of \CPSs. We empirically demonstrated that the \BO method is compatible with state-of-the-art testing approaches.

Discussion of Optimization Methods: The performance of each evaluated \BO method depends on the \SUT and how the falsified area or falsified points are located based on the input ranges for each benchmark problem. Vanilla BO, which is the standard global \BO method, showed a good performance for those specifications and problems that can be falsified at boundaries of input spaces or corners. When the number of dimensions is high \VanillaBO searches the edges more because the prediction error and the uncertainty of the surrogate model at the edges become large. On the other hand, using local regions, exploration and exploitation in \turbo are limited to local trust regions, which results in less efficient falsification compared to \VanillaBO for those easily falsifiable problems where failure points are at corners. Hence, \VanillaBO showed better performance for those benchmark problems with the ability to be falsified at corners.
% Zahra: Is the following sentence about Turbo true? Line 149.
However, the efficiency of \turbo increases with the number of dimensions showing a remarkable performance even for hard problems that are not falsified with other methods. Compared to the other \BO methods for high-dimensional settings, such as the previously proposed REMBO~\cite{Wang2016REMBO}, \turbo does not need any setup, it can be used as an out-of-box tool. The $\pi$BO allows injecting prior knowledge about a failure region. $\pi$BO has shown good performance for those problems that are easily falsified at corners due to the U-shaped distribution that was used as prior knowledge based on previous falsification experiments. Even though $\pi$BO does not scale well in a high dimensional setting, it still shows good performance for a moderate number of dimensions of the input space. 

%In particular, it has 62.87\% success rate for those hard examples in Table~\ref{tab:Hard_to_Falsify} with the input dimensionality less than 40 dimensions. 

Discussion of Acquisition Functions:
% Comparison among the acquisition functions
Comparing the performance of different acquisition functions in \turbo, \LCB showed the best performance. \LCB depends on mean and variance, not on the target value, as \PI does. A challenging aspect of the \LCB is choosing the optimal value for $\beta$. \TS, on the contrary, does not require any parameters to be selected. Choosing the best target value for \PI is difficult as it depends on the application and knowing where the optimal value is. This information is, in general, unknown to us. While the default setting in \BO is to use the best-current value, we here assume constant values of $0$ and $-1$ to emphasize failure points. For those specifications with minimum optima between $(-1, 0)$, \PI does not perform well because it assumes that $-1$ is the lowest objective value and neglects the failure points having an objective value $>-1$ from the surrogate model, that indeed falsifies the specification. 
% TS discussion
Objective function values calculated using the \maxS semantics can be constant in large regions. It means the same objective function value for different input parameters. Hence, optimization methods cannot get any sense of direction from the objective function. A benefit of using \TS as an acquisition function is its randomness which may detect falsified points where there is no information from the objective function to guide the optimization process. In \TS, once we have a trained surrogate model, the concept is to greedily sample a configuration from the posterior
with the lowest value, and sampling from the posterior generates TS’s randomness.
As \TS is focused more on exploitation, it is more efficient w.r.t. other acquisition functions.

% Conclusion
\section{Conclusion}
\label{Sec:Conclusion}

In this study, we assess Bayesian Optimization (\BO) for falsifying cyber-physical systems (\CPSs). \BO learns the system under test to strategically choose the next test to execute. We adapt BO in several ways, making it more efficient for falsification compared to the vanilla BO. Specifically, we demonstrate how trust region-based \BO approaches can effectively handle the evaluated falsification problems in this paper with a high number of dimensions, and how prior knowledge can be incorporated into BO. Our work establishes a practical framework for using \BO as an out-of-box tool, unlike previous studies that do not account for its practicality. We propose two notable \BO methods, $\pi$\BO, and \turbo, for falsification and provide a comprehensive evaluation demonstrating their efficiency in solving the standard falsification problems available in the falsification community. In our experiments, \turbo outperforms other state-of-the-art methods for high-dimensional and difficult-to-falsify specifications, without requiring any information from the practitioner. $\pi$BO allows the injection of prior knowledge about falsification if available. For benchmark examples with fewer input parameters for optimization, $\pi$BO performs as well as or even better than \hybrid when the specifications are falsifiable at corners or boundaries. A correct prior about falsification, such as a U-shaped distribution emphasizing corners, can increase efficiency and reduce the number of simulations needed for falsification. Additionally, we conduct a comprehensive evaluation of different acquisition functions in \turbo, propose modifications for falsification, and determine that using a \LCB in \turbo is the optimal choice, as it is less challenging to set up the parameters for good performance on evaluated benchmark examples. In future work, we plan to combine various \BO methods to further enhance the overall efficiency.

\section*{Acknowledgments}
This work was supported by the Swedish Research Council (VR) project
SyTeC VR 2016-06204 and from the Swedish Governmental Agency for
Innovation Systems (VINNOVA) under project TESTRON 2015-04893, and was partly supported by the Wallenberg AI, Autonomous Systems and Software program (WASP) funded by the Knut and Alice Wallenberg Foundation. This research was also supported in part by affiliate members and other supporters of the Stanford DAWN project—Ant Financial, Facebook, Google, InfoSys, Teradata, NEC, and VMware. The evaluations were performed using resources at High Performance Computing Center North (HPC2N), Umeå University, a Swedish national center for Scientific and Parallel Computing.

The authors would like to thank Martin Fabian, Koen Claessen, and Nicholas Smallbone, for their helpful
comments on this paper.

%Bibliography
\bibliographystyle{unsrt}  
\bibliography{references.bib}  

\begin{thebibliography}{10}

\bibitem{alur2015principles}
Rajeev Alur.
\newblock {\em Principles of cyber-physical systems}.
\newblock MIT press, 2015.

\bibitem{audet2017}
Charles Audet and Warren Hare.
\newblock {\em Derivative-free and blackbox optimization}.
\newblock Springer Series in Operations Research and Financial Engineering.
  Springer, 2017.

\bibitem{Hooke1961direct}
Robert Hooke and Terry~A Jeeves.
\newblock `{D}irect search' solution of numerical and statistical problems.
\newblock {\em Journal of the ACM (JACM)}, 8(2):212--229, 1961.

\bibitem{Shahriari2016}
B.~{Shahriari}, K.~{Swersky}, Z.~{Wang}, R.~P. {Adams}, and N.~{de Freitas}.
\newblock Taking the human out of the loop: A review of {B}ayesian
  optimization.
\newblock {\em Proceedings of the IEEE}, 104(1):148--175, 2016.

\bibitem{hvarfner2022pi}
Carl Hvarfner, Danny Stoll, Artur Souza, Marius Lindauer, Frank Hutter, and
  Luigi Nardi.
\newblock $\pi${BO}: Augmenting acquisition functions with user beliefs for
  {B}ayesian optimization.
\newblock {\em arXiv preprint arXiv:2204.11051}, 2022.

\bibitem{vsehic2021lassobench}
Kenan {\v{S}}ehi{\'c}, Alexandre Gramfort, Joseph Salmon, and Luigi Nardi.
\newblock Lassobench: A high-dimensional hyperparameter optimization benchmark
  suite for lasso.
\newblock {\em arXiv preprint arXiv:2111.02790}, 2021.

\bibitem{mayr2022skill}
Matthias Mayr, Faseeh Ahmad, Konstantinos Chatzilygeroudis, Luigi Nardi, and
  Volker Krueger.
\newblock Skill-based multi-objective reinforcement learning of industrial
  robot tasks with planning and knowledge integration.
\newblock {\em arXiv preprint arXiv:2203.10033}, 2022.

\bibitem{berkenkamp2021bayesian}
Felix Berkenkamp, Andreas Krause, and Angela~P Schoellig.
\newblock Bayesian optimization with safety constraints: safe and automatic
  parameter tuning in robotics.
\newblock {\em Machine Learning}, pages 1--35, 2021.

\bibitem{Deshmukh2017Bayesian}
Jyotirmoy Deshmukh, Marko Horvat, Xiaoqing Jin, Rupak Majumdar, and Vinayak~S.
  Prabhu.
\newblock Testing cyber-physical systems through {B}ayesian optimization.
\newblock {\em ACM Trans. Embed. Comput. Syst.}, 16(5s), sep 2017.

\bibitem{hansen2006cma}
Nikolaus Hansen.
\newblock The {CMA} evolution strategy: a comparing review.
\newblock {\em Towards a new evolutionary computation}, pages 75--102, 2006.

\bibitem{Wang2016REMBO}
Ziyu Wang, Frank Hutter, Masrour Zoghi, David Matheson, and Nando de~Feitas.
\newblock {B}ayesian optimization in a billion dimensions via random
  embeddings.
\newblock {\em Journal of Artificial Intelligence Research}, 55:361--387, 2016.

\bibitem{nayebi2019}
A.~Nayebi, A.~Munteanu, and M.~Poloczek.
\newblock A framework for {B}ayesian optimization in embedded subspaces.
\newblock In {\em Proceedings of the 36th International Conference on Machine
  Learning}, page 4752–4761, 2019.

\bibitem{alebo}
B.~Letham, R.~Calandra, A.~Rai, and E.~Bakshy.
\newblock Re-examining linear embeddings for high-dimensional {B}ayesian
  optimization.
\newblock In {\em Advances in Neural Information Processing Systems 33 (NeurIPS
  2020)}, volume~33, pages 1546--1558, 2020.

\bibitem{Ernst2019-ARCH19:}
Gidon Ernst, Paolo Arcaini, Alexandre Donz{\'e}, Georgios Fainekos, Logan
  Mathesen, Giulia Pedrielli, Shakiba Yaghoubi, Yoriyuki Yamagata, and Zhenya
  Zhang.
\newblock {ARCH-COMP} 2019 category report: Falsification.
\newblock In {\em ARCH19. 6th International Workshop on Applied Verification of
  Continuous and Hybrid Systems}, volume~61, pages 129--140. EasyChair, 2019.

\bibitem{Ramezani2020-wodes}
Zahra Ramezani, Johan~Lid{\'e}n Eddeland, Koen Claessen, Martin Fabian, and
  Knut {\AA}kesson.
\newblock Multiple objective functions for falsification of cyber-physical
  systems.
\newblock {\em IFAC-PapersOnLine}, 53(4):417--422, 2020.

\bibitem{Eriksson2019turbo}
David Eriksson, Michael Pearce, Jacob Gardner, Ryan~D Turner, and Matthias
  Poloczek.
\newblock Scalable global optimization via local {B}ayesian optimization.
\newblock {\em Advances in Neural Information Processing Systems},
  32:5496--5507, 2019.

\bibitem{Yuan2000ReviewTrust}
Ya-xiang Yuan.
\newblock A review of trust region algorithms for optimization.
\newblock In {\em Iciam}, volume 99 (1), pages 271--282, 2000.

\bibitem{Thompson1933likelihood}
William~R Thompson.
\newblock On the likelihood that one unknown probability exceeds another in
  view of the evidence of two samples.
\newblock {\em Biometrika}, 25(3/4):285--294, 1933.

\bibitem{Srinivas2010UCB}
Niranjan Srinivas, Andreas Krause, Sham Kakade, and Matthias Seeger.
\newblock Gaussian process optimization in the bandit setting: No regret and
  experimental design.
\newblock In {\em Proceedings of the 27th International Conference on
  International Conference on Machine Learning}, ICML'10, page 1015–1022,
  Madison, WI, USA, 2010. Omnipress.

\bibitem{Kushner1964PI}
Harold~J. Kushner.
\newblock A new method of locating the maximum point of an arbitrary multipeak
  curve in the presence of noise.
\newblock {\em Journal of Basic Engineering}, 86:97--106, 1964.

\bibitem{Ramezani2021_Line}
Zahra Ramezani, Koen Claessen, Nicholas Smallbone, Martin Fabian, and Knut
  Åkesson.
\newblock Testing cyber–physical systems using a line-search falsification
  method.
\newblock {\em IEEE Transactions on Computer-Aided Design of Integrated
  Circuits and Systems}, 41(8):2393--2406, 2022.

\bibitem{Souza2021Prior}
Artur Souza, Luigi Nardi, Leonardo~B Oliveira, Kunle Olukotun, Marius Lindauer,
  and Frank Hutter.
\newblock {B}ayesian optimization with a prior for the optimum.
\newblock In {\em Joint European Conference on Machine Learning and Knowledge
  Discovery in Databases}, pages 265--296. Springer, 2021.

\bibitem{Eddeland2020-ARCH}
Johan~Lid{\'e}n Eddeland, Sajed Miremadi, and Knut {\AA}kesson.
\newblock Evaluating optimization solvers and robust semantics for
  simulation-based falsification.
\newblock {\em EPiC Series in Computing}, 74:259--266, 2020.

\bibitem{Nelder1965}
J.~A. Nelder and R.~Mead.
\newblock {A Simplex Method for Function Minimization}.
\newblock {\em The Computer Journal}, 7(4):308--313, 01 1965.

\bibitem{huyer2008snobfit}
Waltraud Huyer and Arnold Neumaier.
\newblock {SNOBFIT}--stable noisy optimization by branch and fit.
\newblock {\em ACM Transactions on Mathematical Software (TOMS)}, 35(2):1--25,
  2008.

\bibitem{romeijn1994simulated}
H~Edwin Romeijn and Robert~L Smith.
\newblock Simulated annealing for constrained global optimization.
\newblock {\em Journal of Global Optimization}, 5:101--126, 1994.

\bibitem{Mathesen2021}
Logan Mathesen, Giulia Pedrielli, and Georgios Fainekos.
\newblock Efficient optimization-based falsification of cyber-physical systems
  with multiple conjunctive requirements.
\newblock In {\em 2021 IEEE 17th International Conference on Automation Science
  and Engineering (CASE)}, pages 732--737, 2021.

\bibitem{Akazaki2016falsification}
Takumi Akazaki.
\newblock Falsification of conditional safety properties for cyber-physical
  systems with gaussian process regression.
\newblock In {\em International Conference on Runtime Verification}, pages
  439--446. Springer, 2016.

\bibitem{Silvetti2017active}
Simone Silvetti, Alberto Policriti, and Luca Bortolussi.
\newblock An active learning approach to the falsification of black box
  cyber-physical systems.
\newblock In {\em International Conference on Integrated Formal Methods}, pages
  3--17. Springer, 2017.

\bibitem{Claessen2018}
Koen Claessen, Nicholas Smallbone, Johan Eddeland, Zahra Ramezani, and Knut
  {\AA}kesson.
\newblock Using valued booleans to find simpler counterexamples in random
  testing of cyber-physical systems.
\newblock {\em IFAC-PapersOnLine}, 51(7):408--415, 2018.

\bibitem{Donze2010b}
Alexandre Donz{\'e} and Oded Maler.
\newblock Robust satisfaction of temporal logic over real-valued signals.
\newblock In Krishnendu Chatterjee and Thomas~A. Henzinger, editors, {\em
  Formal Modeling and Analysis of Timed Systems}, pages 92--106, Berlin,
  Heidelberg, 2010. Springer Berlin Heidelberg.

\bibitem{Fainekos2009}
Georgios~E. Fainekos and George~J. Pappas.
\newblock Robustness of temporal logic specifications for continuous-time
  signals.
\newblock {\em Theoretical Computer Science}, 410(42):4262 -- 4291, 2009.

\bibitem{Maler2004}
Oded Maler and Dejan Nickovic.
\newblock Monitoring temporal properties of continuous signals.
\newblock In {\em Formal Techniques, Modelling and Analysis of Timed and
  Fault-Tolerant Systems}, pages 152--166, Hei2004.

\bibitem{rasmussen}
CE. Rasmussen and CKI. Williams.
\newblock {\em Gaussian Processes for Machine Learning}.
\newblock Adaptive Computation and Machine Learning. MIT Press, 2006.

\bibitem{frazier2018}
P.~I. Frazier.
\newblock A tutorial on {B}ayesian optimization.
\newblock {\em arXiv preprint arXiv:1807.02811}, 2018.

\bibitem{Nardi2019practical}
Luigi Nardi, David Koeplinger, and Kunle Olukotun.
\newblock Practical design space exploration.
\newblock In {\em 2019 IEEE 27th International Symposium on Modeling, Analysis,
  and Simulation of Computer and Telecommunication Systems (MASCOTS)}, pages
  347--358. IEEE, 2019.

\bibitem{Echard2011ak}
Benjamin Echard, Nicolas Gayton, and Maurice Lemaire.
\newblock {AK-MCS}: an active learning reliability method combining kriging and
  monte carlo simulation.
\newblock {\em Structural Safety}, 33(2):145--154, 2011.

\bibitem{Schobi2017rare}
Roland Sch{\"o}bi, Bruno Sudret, and Stefano Marelli.
\newblock Rare event estimation using polynomial-chaos kriging.
\newblock {\em ASCE-ASME Journal of Risk and Uncertainty in Engineering
  Systems, Part A: Civil Engineering}, 3(2):D4016002, 2017.

\bibitem{Donze2010}
Alexandre Donz{\'e}.
\newblock Breach, a toolbox for verification and parameter synthesis of hybrid
  systems.
\newblock In Tayssir Touili, Byron Cook, and Paul Jackson, editors, {\em
  Computer Aided Verification}, pages 167--170. Springer Berlin Heidelberg,
  2010.

\bibitem{Hoxha2014BenchmarksFT}
Bardh Hoxha, Houssam Abbas, and Georgios Fainekos.
\newblock Benchmarks for temporal logic requirements for automotive systems.
\newblock In {\em ARCH@CPSWeek}, 2014.

\bibitem{Hu2000}
Jianghai Hu, John Lygeros, and Shankar Sastry.
\newblock Towards a theory of stochastic hybrid systems.
\newblock In Nancy Lynch and Bruce~H. Krogh, editors, {\em Hybrid Systems:
  Computation and Control}, pages 160--173. Springer Berlin Heidelberg, 2000.

\bibitem{Dokhanchi2018ARCHCOMP18CR}
Adel Dokhanchi, Shakiba Yaghoubi, Bardh Hoxha, Georgios Fainekos, Gidon Ernst,
  Zhenya Zhang, Paolo Arcaini, Ichiro Hasuo, and Sean Sedwards.
\newblock Arch-comp18 category report: Results on the falsification benchmarks.
\newblock In {\em ARCH@ADHS}, 2018.

\bibitem{MATLAB:2020}
MathWorks.
\newblock Design {NARMA-L2} {N}eural {C}ontroller in {S}imulink.
\newblock
  \url{https://au.mathworks.com/help/deeplearning/ug/design-narma-l2-neural-controller-in-simulink.html},
  2020.
\newblock {Online: accessed 1 March 2021}.

\bibitem{Frehse2018}
Goran Frehse, Alessandro Abate, Dieky Adzkiya, Lei Bu, Mirco Giacobbe,
  Muhammad~Syifa'Ul Mufid, and Enea Zaffanella.
\newblock Arch-comp18 category report: Hybrid systems with piecewise constant
  dynamics.
\newblock In {\em ARCH18. 5th International Workshop on Applied Verification of
  Continuous and Hybrid Systems}, volume~54, pages 1--13. EasyChair, 2018.

\bibitem{Yaghoubi2018}
Shakiba Yaghoubi and Georgios Fainekos.
\newblock Gray-box adversarial testing for control systems with machine
  learning components.
\newblock In {\em Proceedings of the 22nd ACM International Conference on
  Hybrid Systems: Computation and Control}, HSCC '19, page 179–184, 2019.

\bibitem{ARCH16:Hybrid_Modelling_of_Wind2017}
Simone Schuler, Fabiano~Daher Adegas, and Adolfo Anta.
\newblock Hybrid modelling of a wind turbine.
\newblock In Goran Frehse and Matthias Althoff, editors, {\em ARCH16. 3rd
  International Workshop on Applied Verification for Continuous and Hybrid
  Systems}, volume~43 of {\em EPiC Series in Computing}, pages 18--26.
  EasyChair, 2017.

\bibitem{Jin2014Powertrain}
Xiaoqing Jin, Jyotirmoy~V. Deshmukh, James Kapinski, Koichi Ueda, and Ken
  Butts.
\newblock Powertrain control verification benchmark.
\newblock In {\em Proceedings of the 17th International Conference on Hybrid
  Systems: Computation and Control}, HSCC ’14, page 253–262. Association
  for Computing Machinery, 2014.

\bibitem{dang2004verification}
Thao Dang, Alexandre Donz{\'e}, and Oded Maler.
\newblock Verification of analog and mixed-signal circuits using hybrid system
  techniques.
\newblock In {\em International Conference on Formal Methods in Computer-Aided
  Design}, pages 21--36. Springer, 2004.

\bibitem{dokhanchi2015requirements}
Adel Dokhanchi, Aditya Zutshi, Rahul~T Sriniva, Sriram Sankaranarayanan, and
  Georgios Fainekos.
\newblock Requirements driven falsification with coverage metrics.
\newblock In {\em Proceedings of the 12th International Conference on Embedded
  Software}, pages 31--40, 2015.

\bibitem{Yaghoubi:SC}
Shakiba Yaghoubi and Georgios Fainekos.
\newblock Gray-box adversarial testing for control systems with machine
  learning components.
\newblock In {\em Proceedings of the 22nd ACM International Conference on
  Hybrid Systems: Computation and Control}, pages 179--184, 2019.

\bibitem{RamezaniARCH21Pulse}
Zahra Ramezani, Alexandre Donze, Martin Fabian, and Knut {\AA}kesson.
\newblock Temporal logic falsification of cyber-physical systems using input
  pulse generators.
\newblock In Goran Frehse and Matthias Althoff, editors, {\em 8th International
  Workshop on Applied Verification of Continuous and Hybrid Systems (ARCH21)},
  volume~80 of {\em EPiC Series in Computing}, pages 195--202. EasyChair, 2021.

\end{thebibliography}

\appendix

% Benchmark Examples

% Benchmark Problems:
\section{Benchmark Problems}
\label{Sec:Benchmark_Examples}

% Which benchmark problems we work with:
All evaluated benchmark problems~\cite{Ernst2019-ARCH19:, Ramezani2020-wodes} are introduced here briefly. Table~\ref{tab:specifications} presents the STL specifications for all these benchmark problems. 

% All specifications:
\begin{table}[!htbp] \scriptsize
\renewcommand{\arraystretch}{1.5}
\caption{Specifications to falsify for all benchmark problems}
\label{tab:specifications}
\centering
%\begin{adjustbox}{width=0.6\textwidth}
\begin{tabular}{|c|c|c|}

\hline
Spec. & Formula\\
\hline
\hline
$\varphi_1^{AT}$ & $\Box_{[0, 20]}(v < 120)$\\
\hline
$\varphi_2^{AT}$ & $\Box_{[0, 10]}(\omega < 4750)$\\
\hline
$\varphi_3^{AT}$ & $\Box_{[0, 30]}\Big (\big(\lnot g_1 \land \circ~g_1 \big) \implies \circ~\Box_{[0, 2.5]} g_1 \Big) $\\
\hline
$\varphi_4^{AT}$ & $\Box_{[0, 30]}\Big (\big(\lnot g_2 \land \circ~g_2 \big) \implies \circ~\Box_{[0, 2.5]} g_2 \Big) $\\
\hline
$\varphi_5^{AT}$ & $\Box_{[0, 30]}\Big (\big(\lnot g_3 \land \circ~g_3 \big) \implies \circ~\Box_{[0, 2.5]} g_3 \Big) $\\
\hline
$\varphi_6^{AT}$ & $\Box_{[0, 30]}\Big (\big(\lnot g_4 \land \circ~g_4 \big) \implies \circ~\Box_{[0, 2.5]} g_4 \Big) $\\
\hline
$\varphi_7^{AT}$ & $ \big (\Box_{[0, 30]} \omega < 3000 \big) \implies \big (\Box_{[0, 4]} v < 35 \big) $\\
\hline 
$\varphi_8^{AT}$ & $ \big (\Box_{[0, 30]} \omega < 3000 \big) \implies \big (\Box_{[0, 8]} v < 50 \big) $\\
\hline 
$\varphi_9^{AT}$ & $ \big (\Box_{[0, 30]} \omega < 3000 \big) \implies \big (\Box_{[0, 20]} v < 65 \big) $\\
\hline
%\hline
%$\varphi^{\Delta - \Sigma}$ & $\Box\left(\bigwedge_{i=1}^3 (-1 \leq x_i \land x_i \leq 1 )\right).$\\
%\hline
%\hline
%$\varphi^{SS}$ & $\Box(y \geq 0)$\\
%\hline
\hline
$\varphi^{AFC}_1$ & $\Box_{[11,50]} \Big (((\theta < 8.8) \land (\lozenge_{[0,0.05]} (\theta > 40))$ 
\\
& $\lor  (\theta > 40) \land (\lozenge_{[0,0.05]} (\theta < 8.8)) \implies (\Box_{[1,5]} |\mu| < 0.008) \Big) $\\
\hline
$\varphi^{AFC}_2$ & $\Box_{[11,50]} |\mu| < 0.007 $\\
\hline
\hline
$\varphi^{NN}_1$ & $\Box_{[1,37]} \Big (\neg (|\mathit{Pos} - \mathit{Ref}| > 0.005 + 0.03 |Ref|)$ 
\\
& $\implies \lozenge_{[0,2]} \Box_{[0,1]} \big(  0.005 + 0.03 |Ref| \leq |Pos-Ref| \big) \Big )$ \\
\hline
\hline
$\varphi^{NN}_2$ & $\Box_{[1,37]} \Big  (\neg (|Pos -Ref| > 0.005 + 0.04 |Ref|)$ 
\\
& $\implies \lozenge_{[0,2]} \Box_{[0,1]} \big(  0.005 + 0.04 |Ref| \leq |Pos-Ref| \big) \Big )$ \\
\hline
\hline
$\varphi^{WT}_1$ & $\Box_{[30,630]} \theta \leq 14.2$ \\
\hline
$\varphi^{WT}_2$ & $\Box_{[30,630]} 21000 \leq M_{g,d} \leq 47500$ \\
\hline
$\varphi^{WT}_3$ & $\Box_{[30,630]} \Omega \leq 14.3$ \\
\hline
$\varphi^{WT}_4$ & $\Box_{[30,630]} \lozenge_{[0,5]} |\theta - \theta_{d}| \leq 1.6$ \\
\hline
\hline
$\varphi^{CC}_1$ & $\Box_{[0,100]} (y_5 - y_4 \leq 40$) \\
\hline
$\varphi^{CC}_2$ & $\Box_{[0,100]} \lozenge_{[0,30]} y_5 - y_4 \geq 15$ \\
\hline
$\varphi^{CC}_3$ & $\Box_{[0,80]} \Big ((\Box_{[0,20]} y_2-y_1 \leq 20) \lor (\lozenge_{[0,20]} y_5 - y_4 \geq 40) \Big) $ \\
\hline
$\varphi^{CC}_4$ & $\Box_{[0,65]} \lozenge_{[0,30]} \Box_{[0,20]} (y_5 - y_4 \geq 8)$ \\
\hline
$\varphi^{CC}_5$ & $\Box_{[0,72]} \lozenge_{[0,8]} \Big ( (\Box_{[0,5]} y_2 - y_1 \geq 9) \implies (\Box_{[5,20]} y_5 - y_4 \geq 9) \Big)$ \\
\hline
\hline
$\varphi^{F_{16}}$ & $\Box_{[0,15]} altitude > 0$ \\
\hline
\hline
$\varphi^{SC}$ & $\Box_{[30,35]} \Big (87 \leq pressure \land pressure \leq 87.5 \Big ) $ \\
\hline
\hline
$\varphi_1^{AT^\prime}$ & $\lozenge_{[0, T]}(\omega \geq 2000)$\\
\hline
$\varphi_2^{AT^\prime}$ & $\Box\lozenge_{[0, T]}(\omega \leq 3500 \lor \omega \geq 4500)$\\
\hline
$\varphi_3^{AT^\prime}$ & $\Box_{[0, T]}(\lnot(gear == 4)) $\\
\hline
$\varphi_4^{AT^\prime}$ & $\lozenge(\Box_{[0, T]}(gear == 3))$\\
\hline
$\varphi_5^{AT^\prime}$ & $\bigwedge_{i = 1, \ldots, 4}\Box( (\lnot(gear == i) \land \lozenge_{[0, \epsilon]}(gear == i)$\\
 & $\implies (\Box_{[\epsilon, T + \epsilon]}(gear == i)))$\\
\hline
$\varphi_6^{AT^\prime}$ & $\Box_{[0, T]}(v \leq 85) \lor \lozenge(\omega \geq 4500)$\\
\hline
$\varphi_7^{AT^\prime}$ & $\lnot\Big((\Box_{[0,1]}gear == 1) \land (\Box_{[2,4]}gear == 2) $ \\
 & $\land(\Box_{[5,7]}gear == 3) \land (\Box_{[8,10]}gear == 3) $ \\
  & $\land(\Box_{[12,15]}gear == 2)\Big)$ \\
\hline 
$\varphi_8^{AT^\prime}$ & $\Box_{[0,20]}\big((gear==4 \land throttle > 45 $\\
 & $\land throttle < 50) \implies \omega < \bar{\omega}\big)$\\
\hline
\hline
$\varphi^{\Delta - \Sigma}$ & $\Box\left(\bigwedge_{i=1}^3 (-1 \leq x_i \land x_i \leq 1 )\right).$\\
\hline
\hline
$\varphi^{SS}$ & $\Box(y \geq 0)$\\
\hline
\end{tabular}
%\end{adjustbox}
\end{table}

% AT
\paragraph{\textbf{Automatic Transmission (AT)}} 
There are two inputs for this problem, $0 \leq \mathit{throttle} \leq 100$, and $0 \leq \mathit{brake} \leq 325$, which can be active at the same time~\cite{Hoxha2014BenchmarksFT}. This problem has two instances. In Instance~1, both input signals are piecewise constant with a \emph{previous} interpolation between them, corresponds to the previous sample value, while Instance~2 has constrained input signals with discontinuities at most every 5 time units.

% CC
\paragraph{\textbf{Chasing Cars (CC)}} This model has two inputs $0 \leq \mathit{throttle} \leq 1$ and $0 \leq \mathit{brake} \leq 1$~\cite{Hu2000} with two instances. For Instance 1, the input specifications allow any piecewise continuous signals to be distributed equally, with 8 segments. However, in Instance 2, the input signals are piecewise constant signals with a \emph{previous} interpolation with 20 segments.

% AFC:
\paragraph{\textbf{Fuel Control of an Automotive Power Train (AFC)}}
This system has two inputs of $0 \leq \theta \leq 61.1$ and $900 \leq \omega \leq 1100$~\cite{Dokhanchi2018ARCHCOMP18CR}. The input signal $\theta$ is piecewise constant with 10 uniform segments, i.e., a \emph{previous} interpolation while $\omega$ is constant.

% NN:
\paragraph{\textbf{Neural Network Controller (NN)}}
This benchmark problem has a reference value $1 \leq \mathit{Ref} \leq 3$ for the position~\cite{MATLAB:2020} as input. For Instance~1 of this problem, the input signal needs discontinuities to be at least 3 time-units long, 12 segments, while Instance~2 requires exactly 3 constant segments.

% F16:
\paragraph{\textbf{Aircraft Ground Collision Avoidance System (F16)}}
The system~\cite{Frehse2018} is required to start with initial conditions $0.2\pi \leq  roll \leq 0.2833\pi$, $−0.5\pi \leq  pitch \leq  −0.54\pi$, and $0.25\pi \leq yaw \leq  0.375\pi$.

% SC:
\paragraph{\textbf{Steam Condenser with Recurrent Neural Network Controller (SC)}}
The only input of this system is $3.99 \leq  \mathit{Fs} \leq 4.01 $,~\cite{Yaghoubi2018}, and the input signal should be piecewise constant with 12 and 20 evenly spaced segments for Instance~1 and Instance~2, respectively.

% WT:
\paragraph{\textbf{Wind Turbine (WT)}}
A simplified wind turbine model from~\cite{ARCH16:Hybrid_Modelling_of_Wind2017} with only one input $8 \leq v \leq 16$ is considered. The input signal of this problem is piecewise with \emph{spline} interpolation, a cubic polynomial interpolation between the control points, each with specified derivatives.

% AT':
\paragraph{\textbf{Automatic Transmission ($\mathbf{AT^\prime}$)}}
The two inputs to the model are $0 \leq \mathit{throttle}$ $\leq 100$ and $0 \leq \mathit{brake} \leq 500$~\cite{Jin2014Powertrain}. 
This problem has different specifications and a different input range for the $\mathit{brake}$ compared to the ARCH problem presented, $AT$. There are 7 control points for $\mathit{throttle}$ and 3 for $\mathit{brake}$ distributed uniformly with \emph{pchip} interpolation. To distinguish this problem from $AT$, it is called $\mathbf{AT^\prime}$.

% Third Order Modulator:
\paragraph{\textbf{Third Order $\Delta - \Sigma$ Modulator}}

The third order $\Delta - \Sigma$ modulator has one input $U$, three states $x_1, x_2, x_3$, and three initial conditions $x_1^\mathit{init}, x_2^\mathit{init}, x_3^\mathit{init}$, all defined in $[-0.1, 0.1]$~\cite{dang2004verification}. Three different ranges are considered for the input, $-0.35 \leq U \leq 0.35$, $-0.40 \leq U \leq 0.40$, and $-0.45 \leq U \leq 0.45$.

% SS
\paragraph{\textbf{Static Switched (SS)}}
The static switched system is a model without any dynamics inspired by~\cite{dokhanchi2015requirements} with two inputs in the range $[-1, 1]$. Three
different values are considered for parameter $\gamma= 0.7, 0.8, 0.9$.

\section{Other Experiments on ARCH Benchmark}
\label{Other_Examples_Appendix}
The results for the benchmark examples that can be falsified easily with a few simulations regardless of which optimization method are presented in Table~\ref{tab:Instance1}-\ref{tab:Without_Instance}. %and also are included in the tables~\ref{tab:Easy_to_Falsify}-~\ref{tab:Hard_to_Falsify}. 
Table~\ref{tab:Instance1} and \ref{tab:Instance2} refers to result for the specifications of $AT$, $CC$, and $NN$ systems with Instance 1 and Instance 2, respectively. On the other hand, Table~\ref{tab:Without_Instance} includes the results for the specifications of $AT^\prime$, $\varphi_1^{\Delta - \Sigma}$, $WT$ and $AFC$ systems that only one input instance are evaluated on them.

% Instance 1: 
\begin{table}[!ht] \scriptsize
\renewcommand{\arraystretch}{1.5}
\caption{Results for the specifications of $AT$, $CC$, and $NN$ systems, Instance 1 which are not included in the tables~\ref{tab:Easy_to_Falsify}-~\ref{tab:Hard_to_Falsify}.}
\label{tab:Instance1}
\centering
%\begin{tabular}{|c|c|c|c|c|c|c|c|c|c|}
\begin{tabular}{c c c c c c c c c c c}

\toprule
Specifications
& Number of 
& \multicolumn{1}{c}{vanilla} 
& \multicolumn{1}{c}{\turbo} 
& \multicolumn{1}{c}{\turbo} 
& \multicolumn{1}{c}{\turbo}
& \multicolumn{1}{c}{\turbo}
& \multicolumn{1}{c}{$\pi$BO} 
& \multicolumn{1}{c}{LSF} 
& \multicolumn{1}{c}{HCR} 
\\

& Dimensions
& \multicolumn{1}{c}{\BO} 
& \multicolumn{1}{c}{\TS} 
& \multicolumn{1}{c}{\LCB} 
& \multicolumn{1}{c}{\PI ($\tau = 0$)}
& \multicolumn{1}{c}{\PI ($\tau = -1$)}
& \multicolumn{1}{c}{} 
& \multicolumn{1}{c}{} 
& \multicolumn{1}{c}{} 
\\

\toprule

\multirow{10}{*}{}

$\varphi_1^{AT}$
& 8
& 100 (7)
& 100 (86)
& 100 (81)
& 100 (79)
& 100 (80)
& 100 (37)
& 100 (34)
& \textbf{100 (5)}
\\

$\varphi_2^{AT}$
& 8
& \textbf{100 (3)}
& 100 (13)
& 100 (12)
& 100 (9)
& 100 (9)
& 100 (8)
& 100 (9)
& \textbf{100 (3)}
\\

$\varphi_3^{AT}$
& 8
& 100 (70)
& 100 (17)
& 100 (43) 
& 100 (46)
& 100 (22)
& \textbf{100 (12)}
& 100 (37)
& 100 (27)
\\

$\varphi_4^{AT}$
& 8
& 100 (41)
& 100 (16)
& 100 (17)
& 100 (22)
& 100 (14)
& \textbf{100 (11)}
& 100 (31)
& 100 (25)
\\

$\varphi_5^{AT}$
& 8
& 100 (52)
& 100 (22)
& 100 (11)
& 100 (18)
& 100 (18)
& \textbf{100 (11)}
& 100 (21)
& 100 (23)
\\

$\varphi_9^{AT}$
& 8
& 100 (104)
& 100 (28)
& 100 (30)
& \textbf{100 (21)}
& 100 (40)
& 100 (77)
& 100 (22)
& 100 (61)
\\

\midrule

$\varphi_1^{CC}$
& 8
& \textbf{100 (2)}
& 100 (6)
& 100 (9)
& 100 (7)
& 100 (9)
& 100 (4)
& 100 (8)
& 100 (5)
\\

$\varphi_2^{CC}$
& 8
& \textbf{100 (3)}
& \textbf{100 (3)}
& 100 (4)
& 100 (6)
& 100 (5)
& 100 (6)
& 100 (14)
& 100 (5)
\\

$\varphi_3^{CC}$
& 8
& \textbf{100 (2)}
& 100 (13)
& 100 (11)
& 100 (11)
& 100 (12)
& 100 (6)
& 100 (14)
& 100 (5)
\\

$\varphi_5^{CC}$
& 8
& 100 (6)
& 100 (59)
& 100 (55)
& 100 (52)
& 100 (61)
& 100 (18)
& 100 (32)
& \textbf{100 (5)}
\\

\midrule

$\varphi_1^{NN}$
& 12
& 100 (145)
& 100 (32)
& 100 (21)
& 100 (32)
& \textbf{100 (19)}
& 100 (32)
& 100 (25)
& 100 (39)
\\

\bottomrule

\end{tabular}
\end{table}

% Instance 2: 
\begin{table}[!ht] \scriptsize
\renewcommand{\arraystretch}{1.5}
\caption{Results for the specifications of $AT$, $CC$, and $NN$ systems, Instance 2 which are not included in the tables~\ref{tab:Easy_to_Falsify}-~\ref{tab:Hard_to_Falsify}.}
\label{tab:Instance2}
\centering
\begin{tabular}{c c c c c c c c c c c}

\toprule
Specifications
& Number of 
& \multicolumn{1}{c}{vanilla} 
& \multicolumn{1}{c}{\turbo} 
& \multicolumn{1}{c}{\turbo} 
& \multicolumn{1}{c}{\turbo}
& \multicolumn{1}{c}{\turbo}
& \multicolumn{1}{c}{$\pi$BO} 
& \multicolumn{1}{c}{LSF} 
& \multicolumn{1}{c}{HCR} 
\\

& Dimensions
& \multicolumn{1}{c}{\BO} 
& \multicolumn{1}{c}{\TS} 
& \multicolumn{1}{c}{\LCB} 
& \multicolumn{1}{c}{\PI ($\tau = 0$)}
& \multicolumn{1}{c}{\PI ($\tau = -1$)}
& \multicolumn{1}{c}{} 
& \multicolumn{1}{c}{} 
& \multicolumn{1}{c}{} 
\\

\toprule

\multirow{10}{*}{}

$\varphi_3^{AT}$
& 40
& \textbf{100 (2)}
& 100 (5)
& 100 (4)
& 100 (5)
& 100 (4)
& 100 (3)
& 100 (10)
& 100 (7)
\\

$\varphi_4^{AT}$
& 40
& 100 (2)
& \textbf{100 (1)}
& \textbf{100 (1)}
& \textbf{100 (1)}
& 100 (2)
& \textbf{100 (1)}
& 100 (3)
& 100 (3)
\\

$\varphi_5^{AT}$
& 40
& \textbf{100 (1)}
& \textbf{100 (1)}
& \textbf{100 (1)}
& \textbf{100 (1)}
& \textbf{100 (1)}
& \textbf{100 (1)}
& 100 (2)
& 100 (2)
\\

$\varphi_6^{AT}$
& 40
& \textbf{100 (1)}
& 100 (2)
& 100 (2)
& 100 (2)
& \textbf{100 (1)}
& 100 (2)
& 100 (4)
& 100 (3)
\\

\midrule

$\varphi_1^{CC}$
& 40
& 100 (6)
& 100 (59)
& 100 (55)
& 100 (52)
& 100 (61)
& 100 (18)
& 100 (32)
& \textbf{100 (5)}
\\

$\varphi_3^{CC}$
& 40
& \textbf{100 (2)}
& 100 (18)
& 100 (28)
& 100 (24)
& 100 (26)
& 100 (8)
& 100 (13)
& 100 (5)
\\

$\varphi_5^{CC}$
& 40
& 100 (56)
& 100 (49)
& 100 (47)
& 100 (34)
& 100 (70)
& \textbf{100 (31)}
& 100 (67)
& 100 (38)
\\

\midrule

$\varphi_1^{NN}$
& 3
& \textbf{100 (7)}
& 100 (48)
& 100 (37)
& 100 (42)
& 100 (58)
& 100 (11)
& 100 (65)
& 100 (125)
\\

\bottomrule

\end{tabular}
\end{table}

% No Instances: 
\begin{table}[!ht] \scriptsize
\renewcommand{\arraystretch}{1.5}
\caption{Results for the specifications of $AT^\prime$, $\Delta - \Sigma$, $WT$ and $AFC$ systems which are not included in the tables~\ref{tab:Easy_to_Falsify}-~\ref{tab:Hard_to_Falsify}.}
\label{tab:Without_Instance}
\centering
%\begin{tabular}{|c|c|c|c|c|c|c|c|c|c|}
\begin{tabular}{c c c c c c c c c c c}

\toprule
Specifications
& Number of 
& \multicolumn{1}{c}{vanilla} 
& \multicolumn{1}{c}{\turbo} 
& \multicolumn{1}{c}{\turbo} 
& \multicolumn{1}{c}{\turbo}
& \multicolumn{1}{c}{\turbo}
& \multicolumn{1}{c}{$\pi$BO} 
& \multicolumn{1}{c}{LSF} 
& \multicolumn{1}{c}{HCR} 
\\

& Dimensions
& \multicolumn{1}{c}{\BO} 
& \multicolumn{1}{c}{\TS} 
& \multicolumn{1}{c}{\LCB} 
& \multicolumn{1}{c}{\PI ($\tau = 0$)}
& \multicolumn{1}{c}{\PI ($\tau = -1$)}
& \multicolumn{1}{c}{} 
& \multicolumn{1}{c}{} 
& \multicolumn{1}{c}{} 
\\

\toprule

\multirow{10}{*}{}

$\varphi_1^{AT^\prime} (T = 20)$
& 10
& 100 (27)
& 100 (26)
& 100 (27)
& 100 (23)
& 100 (21)
& 90 (82)
& 100 (63)
& \textbf{100 (1)}
\\

$\varphi_1^{AT^\prime} (T = 30)$
& 10
& 100 (73)
& 100 (40)
& 100 (50)
& 100 (43)
& 100 (37)
& 100 (79)
& 100 (172)
& \textbf{100 (1)}
\\

$\varphi_1^{AT^\prime} (T = 40)$
& 10
& 100 (146)
& 95 (128)
& 100 (80)
& 100 (109)
& 100 (60)
& 100 (190)
& 100 (241)
& \textbf{100 (1)}
\\

$\varphi_2^{AT^\prime} (T = 10)$
& 10
& 100 (4)
& 100 (13)
& 100 (16)
& 100 (14)
& 100 (12)
& 100 (11)
& 100 (18)
& \textbf{100 (3)}
\\

$\varphi_3^{AT^\prime} (T = 5)$
& 10
& \textbf{100 (11)}
& 100 (104)
& 100 (60)
& 100 (133)
& 100 (85)
& 100 (36)
& 100 (81)
& 100 (21)
\\

$\varphi_4^{AT^\prime} (T = 2)$
& 10
& 100 (12)
& 100 (22)
& 100 (18)
& 100 (27)
& 100 (29)
& 100 (12)
& 100 (29)
& \textbf{100 (1)}
\\

$\varphi_5^{AT^\prime} (T = 2)$
& 10
& \textbf{100 (1)}
& 100 (4)
& 100 (3)
& 100 (3)
& 100 (3)
& 100 (2)
& 100 (6)
& 100 (10)
\\

$\varphi_8^{AT^\prime} (\bar{\omega} = 3000)$
& 10
& \textbf{100 (4)}
& 100 (9)
& 100 (13)
& 100 (10)
& 100 (7)
& 100 (5)
& 100 (11)
& 100 (5)
\\

\midrule

$\varphi_1^{\Delta - \Sigma}$ $U \in [−0.40, 0.40]$
& 4
& 100 (11)
& 100 (57)
& 100 (79)
& 100 (39)
& 100 (115)
& 100 (14)
& 100 (52)
& \textbf{100 (5)}
\\

$\varphi_1^{\Delta - \Sigma}$ $U \in [−0.45, 0.45]$
& 4
& \textbf{100 (11)}
& 100 (29)
& 100 (54)
& 100 (42)
& 100 (64)
& 100 (21)
& 100 (39)
& \textbf{100 (11)}
\\

\midrule

$\varphi_1^{WT}$
& 126
& \textbf{100 (1)}
& 100 (2)
& \textbf{100 (1)}
& 100 (2)
& \textbf{100 (1)}
& \textbf{100 (1)}
& 100 (3)
& 100 (3)
\\

$\varphi_2^{WT}$
& 126
& \textbf{100 (1)}
& \textbf{100 (1)}
& \textbf{100 (1)}
& \textbf{100 (1)}
& \textbf{100 (1)}
& \textbf{100 (1)}
& 100 (2)
& 100 (2)
\\

$\varphi_3^{WT}$
& 126
& \textbf{100 (1)}
& \textbf{100 (1)}
& \textbf{100 (1)}
& \textbf{100 (1)}
& \textbf{100 (1)}
& \textbf{100 (1)}
& 100 (2)
& 100 (2)
\\

$\varphi_4^{WT}$
& 126
& 100 (37)
& 100 (85)
& 100 (112)
& 100 (125)
& 100 (111)
& 100 (37)
& 100 (66)
& \textbf{100 (30)}
\\

\midrule

$\varphi_1^{AFC}$
& 11
& \textbf{100 (1)}
& 100 (74)
& 100 (81)
& 100 (91)
& 100 (72)
& 100 (24)
& 100 (6)
& 100 (5)
\\

$\varphi_1^{AFC}$
& 11
& \textbf{100 (1)}
& 100 (15)
& 100 (20)
& 100 (15)
& 100 (10)
& 100 (8)
& 100 (2)
& 100 (5)
\\

\bottomrule

\end{tabular}
\end{table}

% Unfalsifiable benchmark examples
\section{Unfalsifiable benchmark examples}
\label{Unfalsifiable_Examples}

For some benchmark examples, falsification is challenging. We show here the result for $\varphi_1^{AT}$ with Instance 2 and $\varphi_1^{SC}$ with both instances in Table~\ref{tab:Not_Falsified}. 
While $\varphi_1^{AT}$, with 8 dimensions, shown in Table~\ref{tab:Instance1}, can be easily falsified, its high-dimensional version with 40 dimensions is hard to falsify, as seen in Table~\ref{tab:Not_Falsified}.
For the \SC problem,~\cite{Yaghoubi:SC} demonstrated that by combining a Simulated Annealing global search with an optimal control-based local search on the infinite-dimensional input space, it is possible to falsify this specification. However, this is not a black-box approach.
In~\cite{RamezaniARCH21Pulse}, the \SC problem is shown to be falsified by using a pulse generator as the input generator. By optimizing over the period, it is possible to find the right period that falsifies this specification.

% No Instances: 
\begin{table}[!htbp] \scriptsize
\renewcommand{\arraystretch}{1.5}
\caption{Results for the specifications of $\varphi_1^{AT}$, Instance 2 and $\varphi_1^{SC}$ for both instances.}
\label{tab:Not_Falsified}
\centering
\begin{tabular}{c c c c c c c c c c c}

\toprule
Specifications
& Instances
& Number of 
& \multicolumn{1}{c}{vanilla} 
& \multicolumn{1}{c}{\turbo} 
& \multicolumn{1}{c}{\turbo} 
& \multicolumn{1}{c}{\turbo}
& \multicolumn{1}{c}{\turbo}
& \multicolumn{1}{c}{$\pi$BO} 
& \multicolumn{1}{c}{LSF} 
& \multicolumn{1}{c}{HCR} 
\\

&
& Dimensions
& \multicolumn{1}{c}{\BO} 
& \multicolumn{1}{c}{\TS} 
& \multicolumn{1}{c}{\LCB} 
& \multicolumn{1}{c}{\PI ($\tau = 0$)}
& \multicolumn{1}{c}{\PI ($\tau = -1$)}
& \multicolumn{1}{c}{} 
& \multicolumn{1}{c}{} 
& \multicolumn{1}{c}{} 
\\

\toprule

\multirow{10}{*}{}

$\varphi_1^{AT}$
&
Instance 2
& 
40
& 0 (-)
& 0 (-)
& 0 (-)
& 0 (-)
& 0 (-)
& 0 (-) 
& 0 (-)
& 0 (-)
\\

\midrule

$\varphi_1^{SC}$
&
Instance 1
&
12
& 0 (-)
& 0 (-)
& 0 (-)
& 0 (-)
& 0 (-)
& 0 (-) 
& 0 (-)
& 0 (-)
\\

$\varphi_1^{SC}$
&
Instance 2
&
20
& 0 (-)
& 0 (-)
& 0 (-)
& 0 (-)
& 0 (-)
& 0 (-) 
& 0 (-)
& 0 (-)
\\

\bottomrule

\end{tabular}
\end{table}

\section{An Aggregated Comparison}

\begin{figure*}[htp]
	\centering
	\centerline{\includegraphics[width=\textwidth]{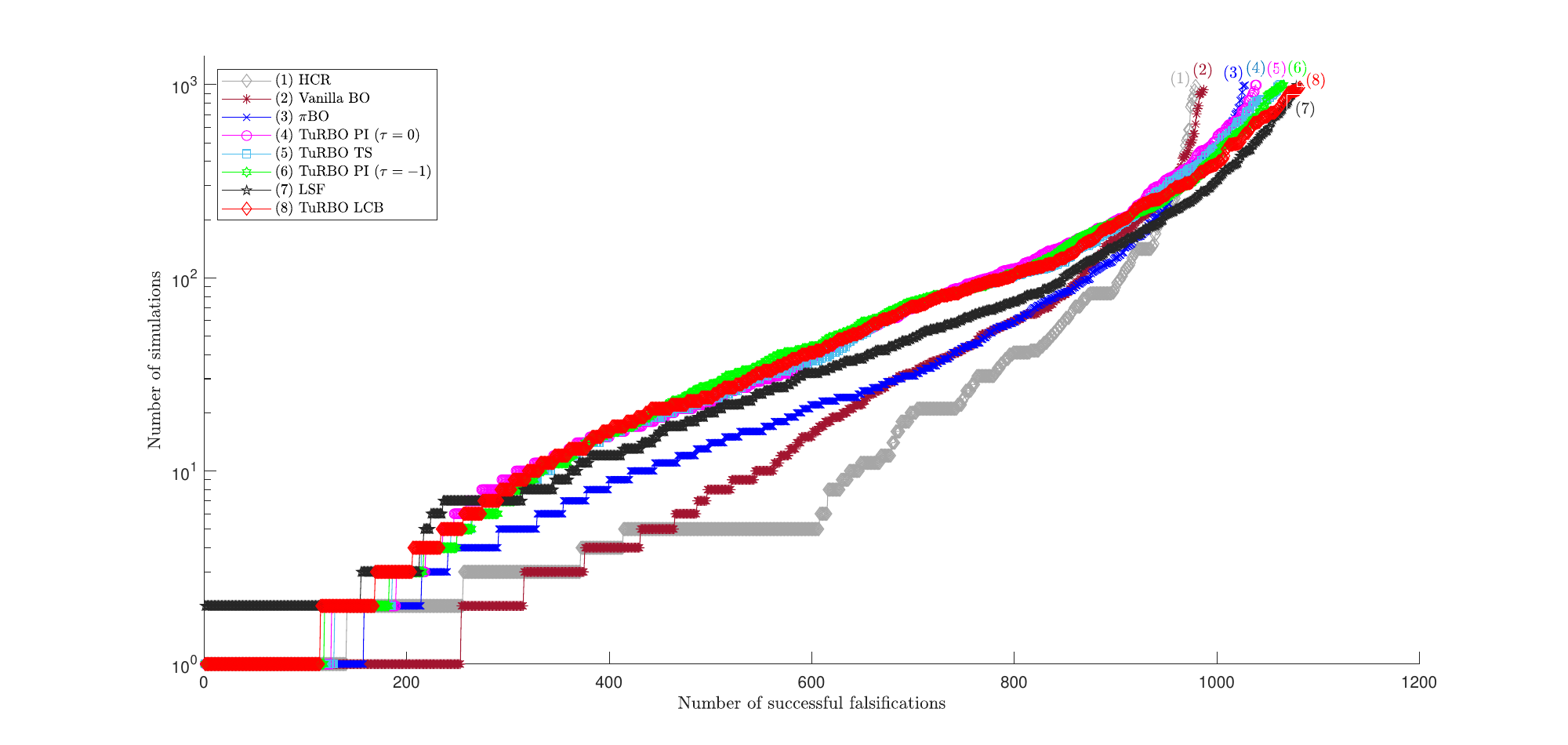}}
	\caption{A cactus plot showing the performance of all examples. The plotted values show how many successful falsifications ($x$-axis) were completed for a given number of simulations ($y$-axis, logarithmic scale). A maximum of 1000 simulations are evaluated.}
	\label{fig:cactus_All_Examples}
\end{figure*}

% Cactus plot for all examples
An aggregated comparison among all methods, including benchmark examples from the appendix, is shown in Fig.~\ref{fig:cactus_All_Examples}. In general, \turbo shows good performance w.r.t. state-of-the-art methods. In particular, selecting \LCB provides better results than \LineSearch. However, \PI and \TS are less efficient than \LineSearch, which was specifically developed to solve falsification problems. 
%Although, \LineSearch still requires doing objective function evaluation directly.
%more expensive evaluations.

\end{document}